%% file: 00.main.tex
\documentclass[sigconf]{acmart}
\acmConference[ASIACCS 2024]{ACM ASIA Conference on Computer and Communications Security}{1-5 July, 2024}{Singapore}

\usepackage{tikz}
\usepackage{algorithmic}
\usepackage{textcomp}
\usepackage{xcolor}
\usepackage{graphicx}
\usepackage{enumitem}
\usepackage{subfig}
\usepackage{bm}
\usepackage{multirow} 
\usepackage{multicol}
\usepackage{flushend}
\usepackage{makecell}
\usepackage{footnote}
\makesavenoteenv{table}
\usepackage{lipsum}
\usepackage{enumitem}
\usepackage{url}

\usepackage[ruled, vlined]{algorithm2e}

\begin{document}

\newcommand{\etal}{\textit{et al.}}
\newcommand{\TN}{BinGo}



\title{\TN{}: Identifying Security Patches in Binary Code with Graph Representation Learning}

% The author information
\author{Xu He} 
\affiliation{%
  \institution{George Mason University}
  \city{Fairfax}
  \state{Virginia}
  \country{USA}
}
\email{xhe6@gmu.edu}

\author{Shu Wang}
\affiliation{%
  \institution{George Mason University}
  \city{Fairfax}
  \state{Virginia}
  \country{USA}}
\email{swang47@gmu.edu}

\author{Pengbin Feng}
\authornote{Pengbin Feng is the corresponding author.}
\affiliation{%
 \institution{Xidian University}
 \city{Xi'an}
 \state{Shanxi}
 \country{China}}
\email{pbfeng@xidian.edu.cn}

\author{Xinda Wang}
\affiliation{%
 \institution{The University of Texas at Dallas}
 \city{Richardson}
 \state{Texas}
 \country{USA}}
\email{xinda.wang@utdallas.edu}

\author{Shiyu Sun}
\affiliation{%
  \institution{George Mason University}
  \city{Fairfax}
  \state{Virginia}
  \country{USA}}
\email{ssun20@gmu.edu}

\author{Qi Li}
\affiliation{%
  \institution{Tsinghua University}
  \city{Beijing}
  % \state{Virginia}
  \country{China}}
\email{qli01@tsinghua.edu.cn}

\author{Kun Sun}
\affiliation{%
  \institution{George Mason University}
  \city{Fairfax}
  \state{Virginia}
  \country{USA}}
\email{ksun3@gmu.edu}

%% Right side of the header
\renewcommand{\shortauthors}{Xu, et al.}

\begin{abstract}
A timely software update is vital to combat the increasing security vulnerabilities. However, some software vendors may secretly patch their vulnerabilities without creating CVE entries or even describing the security issue in their change log. Thus, it is critical to identify these hidden security patches and defeat potential N-day attacks. Researchers have employed various machine learning techniques to identify security patches in open-source software, leveraging the syntax and semantic features of the software changes and commit messages. However, all these solutions cannot be directly applied to the binary code, whose instructions and program flow may dramatically vary due to different compilation configurations. In this paper, we propose BinGo, a new security patch detection system for binary code. The main idea is to present the binary code as code property graphs to enable a comprehensive understanding of program flow and perform a language model over each basic block of binary code to catch the instruction semantics. BinGo consists of four phases, namely, patch data pre-processing, graph extraction, embedding generation, and graph representation learning. Due to the lack of an existing binary security patch dataset, we construct such a dataset by compiling the pre-patch and post-patch source code of the Linux kernel. Our experimental results show BinGo can achieve up to 80.77\% accuracy in identifying security patches between two neighboring versions of binary code. Moreover, BinGo can effectively reduce the false positives and false negatives caused by the different compilers and optimization levels.
\end{abstract}


\maketitle
\input{01.introduction}

\input{02.background}

\input{03.approach}

\input{04.implementation}

\input{05.evaluation}

\input{06.discussion}

\input{07.relatedwork}

\input{08.conclusion}

\begin{acks}
This research is partially supported by ONR grant N00014-23-1-2122, NSF grant CNS-1822094, and NSFC Grant 62132011. 

\end{acks}
% \balance
\bibliographystyle{ACM-Reference-Format}
\bibliography{reference}

\end{document}

%% file: 01.introduction.tex
\section{Introduction}
Software patching is a common practice in software maintenance to ensure the stability, performance, and security of software.
% Patching is the common practice to maintain the software.
Thus, software developers periodically release patch packages to add new features, address performance bugs, and fix security vulnerabilities. Among them, security patches should be prioritized to address software vulnerabilities and thus prevent N-day threats coming from adversaries. However, timely deployment of security patches remains a challenge since not all security patches are explicitly marked~\cite{zhang2021investigation}. For example, software vendors may secretly patch their vulnerabilities without creating CVE entries or even describing the security issue in its change log~\cite{wang2019detecting}. Therefore, identifying these hidden security patches becomes critical for developers and users to improve their software security.

Researchers have leveraged various syntax and semantic features to help identify security patches, such as the software changes as well as their metadata ~\cite{zhou2021spi, wang2020machine, tan2021locating, wu2022enhancing,  wang2021patchrnn, liu2022commitbart, farhi2023detecting, shu2023graphspd, Xu2017spain}. However, those solutions require the availability of source code commits that include source code diffs and commit messages, and unfortunately, not all patches are explicitly released with accessible source code.
% To identify security patches, existing research assumes the availability of patch information (i.e., commit) at the source code level, such as code diff, and commit messages. 
%By leveraging the syntax and semantic features, researchers employed machine learning classifiers to distinguish security patches.
% Researchers constructed syntactic and semantic features from the source code and commit messages, then employed machine learning (ML)~\cite{VulType} or even deep learning (DL) based classifiers to distinguish security patches from other patches.
%clear change logs and 
%
Particularly, commercial closed-source software vendors usually only release a new version of the binary file to replace the old version, presenting challenges in demystifying patch details.

The existing security patch detection methods on source code cannot be directly applied to the binary code due to the limitation of representing the binary code as either a sequential model or a graph model.
The sequential-based solution~\cite{Xu2017spain} treats the changed traces (i.e., a sequence of basic blocks or instructions) as the patch pattern.
% Based on the language models (LMs), e.g., BERT~\cite{devlin2018bert}, GPT~\cite{radford2018improving}, and BART~\cite{lewis2019bart}, sequential-based solutions capture the semantic differences between pre-patch and post-patch code by learning the context between code tokens~\cite{liu2022commitbart, feng2020codebert, li2021palmtree}.
% 
Alternatively, graph-based solutions try to catch the control flow and data dependency by representing source code as code property graphs~\cite{ji2021buggraph, wu2022enhancing, li2019graph, shu2023graphspd, bowman2020vgraph}.
However, the analysis of binary code is more complex than that of source code, since both the instruction sequence and the program flow in binary code may dramatically vary due to different compilation configurations (e.g., compiler versions and optimization levels) used in compiling the source code~\cite{he2022binprov}.
% Besides the sequential statements, 
The sequential-based solutions can hardly accommodate the program flow changes due to the branch and loop statements (which are usually associated with software vulnerabilities~\cite{wang2020machine}) in identifying binary security patches. 
Though recent studies~\cite{wu2022enhancing, shu2023graphspd, bowman2020vgraph}  incorporate graph models to represent program flow, since only statistical features of statements are embedded in each node, they miss subtle code changes in security patches and trigger more false negatives.
In this paper, we propose \TN{}, an end-to-end security patch detection system over binary code. The main idea is to present the code as code property graphs to enable a comprehensive understanding of program flow and perform a Language Model (LM)-based model over each basic block to catch the instruction semantics. \TN{} consists of four phases, namely, patch data pre-processing, graph extraction, embedding generation, and graph representation learning.

\TN{} first identifies and extracts patch-related code segments from a pair of pre-patch and post-patch binaries. Different from the source code analysis that regards code statements as process units, our binary code analysis treats basic blocks of assembly code as process units, which consist of a sequence of contiguous straight-line instructions without branches. Basic blocks are capable of better preserving the inherent code logic than individual assembly instructions, which only perform low-level simple operations, i.e., moving registers. Because we focus on analyzing the code changes between pre-patch and post-patch binaries, it is critical to identify the patch-related basic blocks to reduce the analysis scope. We modify the workflow of DeepBinDiff~\cite{duan2020deepbindiff} to detect patch-related basic blocks with a low false positive rate.
% \TN{} basically follows the technical route of the code feature extraction and then DL-based classification, but we develop novel patch semantic representation and the graph-learning model for detection to sidestep the excessive dependence on LM-based binary analysis solutions.
% avoids excessive dependence on single LP-based semantic representation. 

%our feature extraction
% \TN{} adopts 2 types of representations of patches: a code property graph (CPG) representation and an LM-based embedding representation.
Next, we transform the assembly code into a graph representation to contain  rich syntax and semantic information. We use the code property graph (CPG) that  accommodates the control flow graph (CFG), control dependency graph (CDG), and data dependency graph (DDG) in a unified graph. 
% To embed richer syntax and semantic information of patches, we define a graph representation based on the code property graph (CPG), which accommodates the control flow graph (CFG), control dependency graph (CDG), and data dependency graph (DDG) in a unified graph. 
To reduce the graph size, we implement a graph slicing method to selectively remove context basic blocks that are too distant from the patch-related basic blocks in the control relations. In this step, we generate two code property graphs for pre-patch and post-patch binaries, respectively.
%
%The graph slicing is based on the hypothesis that the removed basic blocks do not have strong relationships with the patch; thus, the removal can reduce noise as well as the graph size.
% By applying a slicing technique, the graph representation covers the patch-related basic blocks and all directly connected basic blocks in CPG, which retain the context of the patch and compress the graph size.
% CPG representation combines the semantic information of the control flow graph (CFG), control dependency graph (CDG), and data dependency graph (DDG).
% Apart from the patch-related basic blocks, we slice the \textit{subcfg} to retain the context of the patch that is 
%
% To better preserve the code semantics, we embed the graphs into a numeric format by using the LLM-based model. 
The graph topology can be represented as an adjacent matrix, while the edge attributes can be embedded according to three relationships, i.e., CFG, CDG, and DDG. 
To fully embed the syntax and semantics in each node (i.e., basic block), we utilize an LM-based model to directly learn an embedding from the assembly instructions. 
% since the language models can effectively capture subtle semantic changes in binary code~\cite{li2021palmtree}. 
Based on the language models, e.g., BERT~\cite{devlin2018bert}, GPT~\cite{radford2018improving}, and BART~\cite{lewis2019bart}, we can capture the real semantic, even if instructions in the node have been changed by compilation~\cite{liu2022commitbart, feng2020codebert, li2021palmtree}.
% We, therefore, generate LM-based embeddings as attributes of nodes in the graph representation, where each node represents a basic block related to the patch.
% In other words, the \textit{subcfg} covers only the patch-related basic blocks and affected blocks.
% For the semantics of the instruction, we leverage the LLM-based embedding model to learn and represent the context semantics and syntax directly from assembly instructions. 
% Existing work shows that the language model can capture subtle semantic changes in binary code. We, therefore, generate LM-based embeddings as attributes of nodes in the graph representation, where each node represents a basic block related to the patch.

% In this paper, we propose \TN, the first security patch detection system in binary code. 
% \TN represents patches by extracting a code property graph (CPG) and conducting the language model (LM) based semantic embedding. 
% CPG combines the semantic information of the control flow graph (CFG), control dependency graph (CDG) and data dependency graph (DDG).
% Besides, we slice the CPG to retain the context of the most relevant basic blocks and compress graph size.
% By applying a slicing technique, we retain the most relevant context and reduce the size of PatchCPG.

Finally, we develop a graph learning model to identify security patches. It is based on the siamese network architecture~\cite{mehrotra2021modeling}, which uses the same weights over both pre-patch and post-patch graphs to obtain a comparable output.
Each branch of the model comprises three graph convolution layers to achieve a detailed understanding of the input graph representation.
% our model
% We then develop a patch-tailored graph neural network (GNN) model for detection, which adopts a siamese structure to capture the difference between patched and unpatched graph representations. 
Since the graphs contain three different types of edges for CFG, CDG, and DDG, we propose a multi-head attention convolution mechanism that views each edge type as an individual convolution channel and aggregates the information of all channels to the next layer.
% Since there are multiple edge attributes including CFG, CDG, and DDG, we propose a new multi-attributed convolution mechanism that views each dimension of edge attributes as an individual convolution channel. 
% The twin networks share weights for patched and unpatched inputs and then concatenate intermediate embeddings into the classifier, which is composed of a fully connected network.  

%our evaluation
We implement a prototype of \TN{} with 2,247 LoC in Python.
% Since there is no existing binary security patch database, we build a benchmark, which is compiled from the existing source code patch dataset, to evaluate \TN{}'s performance. 
Due to the absence of a binary security patch dataset, we first construct a benchmark by compiling the pre-patch and post-patch source code in the Linux kernel. This dataset can be further used in vulnerability detection, patch presence, and hot patch generation. Our experimental results show \TN{} can achieve up to 80.8\% accuracy with an F1 score of 0.76. %, which even outperforms the state-of-the-art approaches in the source code level.
Besides, \TN{} can achieve a false negative ratio of 29.15\% and a false positive ratio of 11.82\%. %which represent a significant step forward in the accurate and efficient identification of security patches.
We further demonstrate \TN{} is effective under different compilation configurations, i.e., compilers and optimization levels, suggesting \TN{} is a viable approach to alleviate the potential false positives caused by compilation factors.

In summary, we make the following contributions:
\vspace{-0.05in}
\begin{itemize}
    % \item We build the Masked LM model to learn assembly code semantics, and then the semantic knowledge can be leveraged to identify binary security patches.
    \item We develop a new binary security patch detection system named \TN{}, which can help users identify potential hidden security patches in newly released binary code and thus prioritize the related system update. 
    % \item We design \TN{}, which builds a graph learning-based Siamese neural network structure to capture differences caused by security patches from patched and unpatched binaries’ assembly code.
    \item We propose a new graph representation for binary code to integrate both the CPG-based static analysis features and the LM-based embedding features, providing a comprehensive representation of subtle code changes in binary code. 
    \item We develop a graph learning-based detection model that adopts a siamese network to identify security patches by comparing the pre-patch and post-patch binaries.
    % \item We propose an automatic approach to generate a security patch dataset according to the pre-collected source code security patch dataset. 
    \item We implement a prototype of \TN{} and evaluate its performance on our benchmark dataset. The experimental results show \TN{} can achieve high accuracy as well as low false positives/negatives on identifying security patches between two neighboring versions of binary code. 
\end{itemize}

%For software developers, security patches among a large number of released patches have the priority for being deployed in the software repositories.

%Binary-level security patches identification could be a good solution to detect N-day and secretly patched vulnerability in released programs.

%We attempt to answer the following research questions in this paper by analyzing vulnerabilities in Linux kernel:
%\begin{itemize}
    %\item How does the binary vulnerabilities and corresponding patches present? (in IR and assembly form)
    %\item How do different levels of compiler optimizations affect the binary patches?
    %\item How do gcc and clang affect the binary patches?
%\end{itemize}

%The main contribution of this paper contains:
%\begin{itemize}
    %\item Aiming at studying the features of security patches for binaries, we build a dataset based on source code patch information collected from NVD. 
    %\item We propose an automatic framework to extract binary-level patch signature via source code patch information.
    %\item Based on the binary dataset, we build an automation detection model based on GNN to identify security patches in released binaries.
    %\item Based on the binary dataset, we summary patched and vulnerable patterns for several vulnerabilities.
    %\item We present a systematic study of both gcc and clang with different optimization levels in compiling vulnerabilities.
%\end{itemize}

% \clearpage

%% file: 02.background.tex
\begin{figure*}[t]
    \centering
    \includegraphics[width=6.9in]{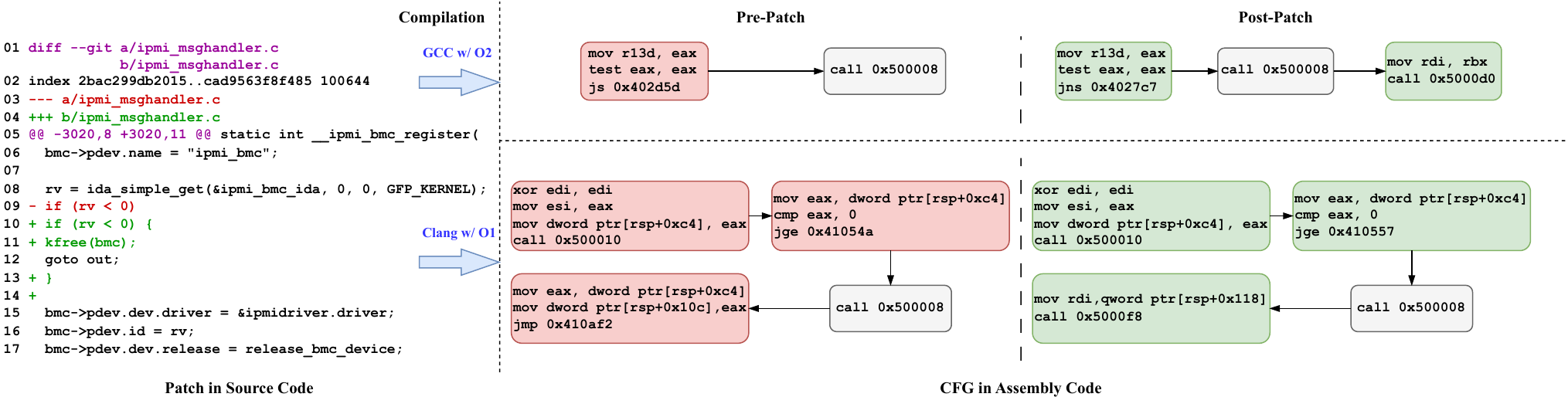}
    % \vspace{-0.05in}
    \caption{A patch in source code and its corresponding assembly code with various compilation configurations (CVE-2019-19046).}
    \label{fig:block_graph}
\end{figure*}

\section{Preliminary}

\subsection{Problem Statement}
% Given two binary versions, we denote the pre-patch version as $bin_a$ and the post-patch version as $bin_b$.
% Compared to $bin_a$, some basic blocks in $bin_b$ have been modified, e.g., adding, deleting, or editing instructions, which can be used to fix vulnerability fix, address performance bugs, or add new functionality. 
% We categorize all code changes into two patch types: the security patch denoted as $p_{s}$ and the non-security patch denoted as $p_{ns}$.
% In this paper, our task is to develop a classification system $f_c$ to check any patch $p$ diffing between $bin_a$ and $bin_b$ is a security patch $p_{s}$ or non-security patch $p_{ns}$. As described in the following formula:

Given two binary versions, we denote the pre-patch version as $bin_a$ and the post-patch version as $bin_b$. Compared to $bin_a$, some basic blocks in $bin_b$ have been modified, e.g., adding, deleting, or editing instructions, which can be used to fix vulnerabilities, address performance bugs, or add new functionalities. We categorize all code changes into two patch types, either the security patch denoted as $p_{s}$ or the non-security patch denoted as $p_{ns}$. In this paper, our task is to develop a classification system $f_c$ to check whether a given patch $p$ is a security patch or a non-security one, as described in Formula (\ref{eq:eq1}).
\vspace{-0.in}
\begin{equation}
    f_c(p_i) = \{p_{s} | p_{ns}\}, w.r.t \{p_0, p_1, \dots,p_n\}\in dif\!f(bin_a,bin_b) 
    \label{eq:eq1}
\end{equation}

% To simplify the identification of security patches, we set two assumptions for the patch presence:
% There is only one patch between pre-patch and post-patch binaries, which means all modified basic blocks found by $diff(bin_a,bin_b)$  belong to only one patch.
% A security patch only fixes one vulnerability, which means all patch-related basic blocks can be connected to a vulnerability by the control or data relationship.
%
% In this paper, we focus on investigating the pre-patch and post-patch code of one commit. 
% We performed a preliminary study on 1,136 commits in the Linux kernel, among which 801 commits contain security fixes. 
% We found 661 commits among them only change code in a single file, while 469 commits even change code in a single function.
% Therefore, we believe that our assumption can cover most scenarios in practice. 

To simplify the security patch identification process, we have two assumptions: (1) There is only one patch between $bin_a$ and $bin_b$, which indicates that all modified basic blocks in $dif\!f(bin_a,bin_b)$ are from the same patch; (2) A security patch only focuses on fixing one vulnerability, which implies that all patch-related basic blocks are connected together via data dependency or control dependency to resolve the vulnerability. We performed a preliminary study on 1,136 patches from the Linux kernel, among which 801 patches are security fixes. There are 661 patches among them only modified one file, and 469 patches only changed one function. Therefore, we believe that our assumption can cover most scenarios in practice.

\subsection{Challenges of Security Patch Detection in Binary Code}
\label{Sec:binaryCode}

Most open-source programs are maintained using Git in practice, which records the code changes before and after the patches~\cite{li2017large}, as shown in the source code part of Figure~\ref{fig:block_graph}. 
Existing work~\cite{wang2020machine} has shown that patches of different purposes can be distinguished based on the corresponding changes in control flow and data flow.
For example, Wang~\etal~found that about 70\% of security patches consist of if-then-else structures~\cite{wang2021PatchDB}, which aim to add or modify the {if} conditions for security check addition or branch statement modification.
However, for non-security patches, the common practice is to add or replace entire procedures (or even functions) to remove redundant code or add new features.
In addition, the code size involved in security patches is usually much smaller than that of non-security patches.
All these uncovered characteristics are very helpful to identify security patches and non-security patches at the source code level.
Existing works~\cite{zhou2021spi,shu2023graphspd} have achieved good recognition performance by conducting multiple graph features and using graph learning algorithms.

However, for binary code, the compiler may introduce extra noise that increases the difficulty of security patch detection.
Due to diverse compiler implementations (e.g., GCC~\cite{GCC} and Clang~\cite{clang}) and optimization mechanisms (e.g., \texttt{O0}, \texttt{O1}, \texttt{O2}, \texttt{O3}, and \texttt{Os}), different binary code variants may be compiled from the same source code, although they execute exactly identical semantics~\cite{he2022binprov,andriesse2016depth}.
As shown in Figure~\ref{fig:block_graph}, we take the patch of CVE-2019-10496 as an example to demonstrate such a difference. 
We compiled the Linux kernel for pre-patch and post-patch versions with two compilation configurations (i.e., Clang with \texttt{O1} and GCC with \texttt{O2}).
We retain the sliced CFG in assembly code, only including the basic blocks of code modification (i.e., the red and green ones) and necessary context. 
The source code shows that the patch merely adds a memory-releasing statement (i.e., \texttt{kfree(bmc)}) in the security check, but such a code change can involve multiple basic blocks in the assembly code. 
Furthermore, we observe that the CFGs derived from Clang and GCC have different instructions, basic blocks, and control dependencies.

With different compilation configurations, the mapping from source code to binary code would be different.
% The mapping of the patch from source code to binary code changes with the compilation configuration.
Compiled by GCC with \texttt{O2}, the source code statement "\texttt{+ kfree(bmc)}" can result in a new basic block and a changed CFG structure in the binary code. 
However, compiled by Clang with \texttt{O1}, the changes only focus on the instructions and there is no effect on basic blocks and CFG structure.
% However, there are only instructions, not basic block and CFG changes in the code compiled by Clang with \texttt{O1}.
Therefore, when detecting security patches in binary code, we also need to consider the additional issues caused by compilation configurations.
In this paper, \TN{} constructs new semantic patterns by fusing assembly instruction embeddings and CPG-based graph representation to incorporate the changes on both code instructions and program flow in the binary patches. 
Moreover, we also build a binary patch dataset based on multiple compilation configurations to evaluate the effectiveness of the \TN{} and its robustness to the compilation.

%% file: 03.approach.tex
\section{System Design}
\label{sec:method}
\begin{figure*}[ht]
    \centering
    \includegraphics[width=6.9in]{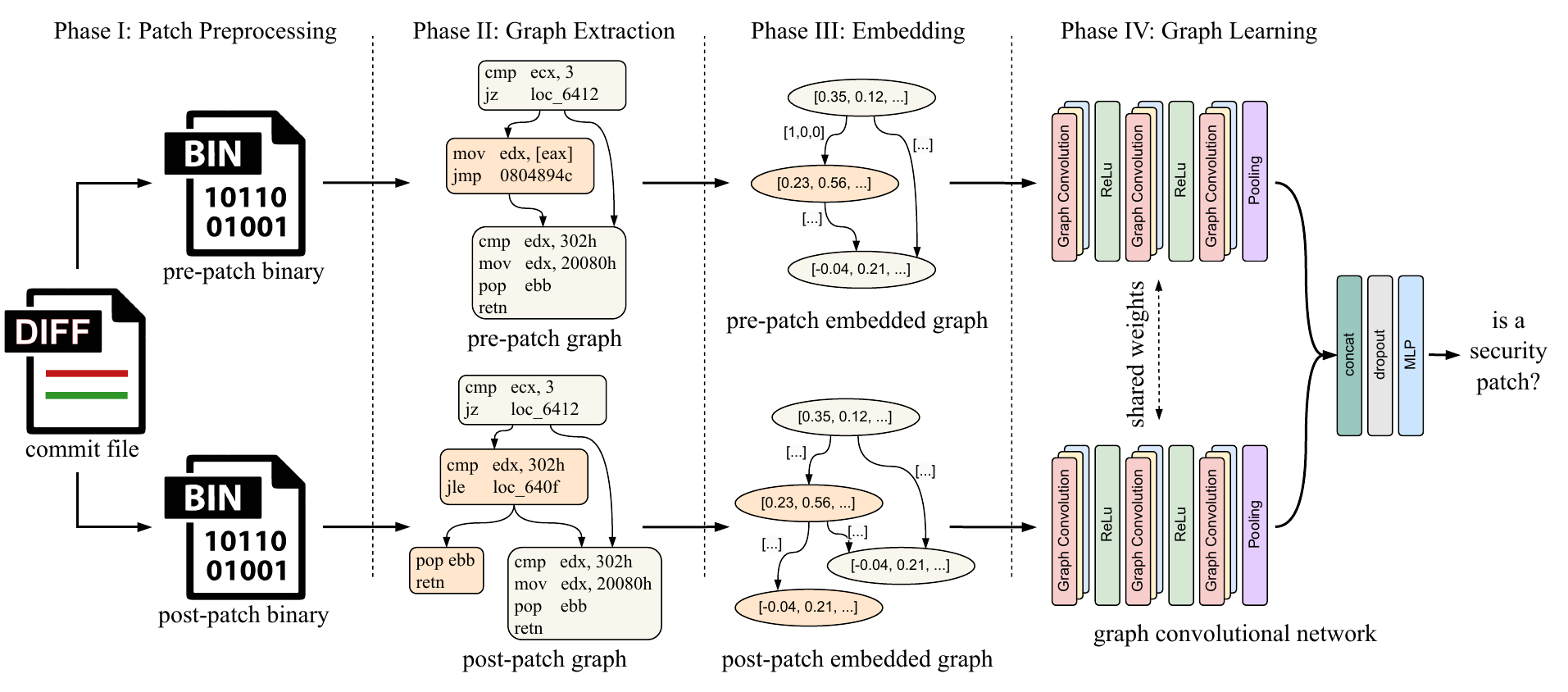}
    \vspace{-0.05in}
    \caption{The overview of the \TN{} system that contains four phases.}
    \label{fig:security_detection}
    % \vspace{-0.05in}
\end{figure*}

% \subsection{System Overview}
\TN{} is an end-to-end deep learning model that detects security patches from non-security ones in binary code. 
Figure~\ref{fig:security_detection} shows the overview of \TN{}, which consists of four phases: patch data pre-processing, graph extraction, embedding generation, and graph representation learning. 
First, \TN{} extracts patch-related code snippets from pre-patch and post-patch binaries.
Second, \TN{} conducts the graph representation by connecting patch-related blocks according to their control flow and data flow dependencies.
Third, \TN{} converts attributes of nodes and edges in the graph into the numeric embedding representations.
Finally, we feed the embeddings into the graph learning model to classify the security patches and the non-security patches.

% from its graph-structured information.
%The goal of the present study is to build an automatic method for binary security patch detection via GNN
%In addition, the generated binary security patch dataset can be used to explore the binary patterns of security patches for patch presence test and 1-day vulnerabilities detection (e.g., caused by secretly released patches).

\vspace{-0.05in}
\subsection{Patch Data Pre-Processing}
\label{Sec:blockextract}
Given a pair of pre-patch and post-patch binary programs, the first phase is to extract the patch-related code segments.
Considering the complex semantic mapping from source code to assembly code during the compilation, modifying one statement in the source code may correspond to one or even multiple basic blocks in the assembly code. 
For instance, the statement "\texttt{c = (a < b) ? a : b;}" can be split into 3 basic blocks. 
Thus, a binary patch can be represented by a set of basic blocks, rather than a single instruction or function.
% After generating binary files, we disassemble patched and unpatched binaries to extract patch-related blocks and sub-graphs.

Formally, we can define the basic block set in pre-patch binary as $BB_{pre}=\{bb_{a_0}, bb_{a_1}, \dots, bb_{a_n}\}$, where $bb_{a_i}$ represents a basic block containing instructions related to the patch.  
To compare the differences before and after the patch, we also define $BB_{post}=\{bb_{b_0}, bb_{b_1}, \dots, bb_{b_m}\}$, which represents all corresponding basic blocks in post-patch binary.

A patch-related block refers to the basic block in which there is at least one changed (i.e., add, delete, or modify) instruction. 
Intuitively, we can detect such blocks by diffing the pre-patch and post-patch disassembled programs.
In practice, we unexpectedly found some rule-based diff tools (e.g., BinDiff in IDA) have high false positives in detecting patch-related blocks~\cite{duan2020deepbindiff}. 
We infer there are two reasons for this issue: 
(1) Basic block changes caused by compilation: the recompilation of modified source code will not only affect the direct-related blocks but also change the neighborhood blocks, especially compiled with a higher optimization level.
(2) The basic blocks of pre-patch and post-patch programs are not in one-to-one correspondence: the modification of a source code statement may correspond to one or even multiple basic blocks.

\begin{figure}[t]
\centering
    \subfloat[Arbitrary binaries.]{
        \includegraphics[width=3.2in]{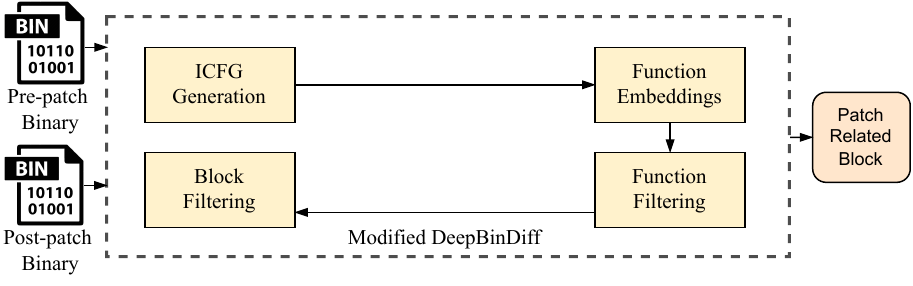}
        }% 
    \hfill
    \subfloat[Images with symbol table.]{
        \includegraphics[width=3.2in]{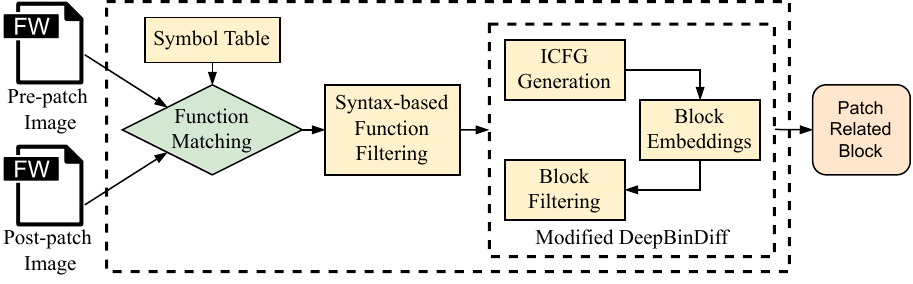}
        }%
  \centering
  \vspace{-0.05in}
  \caption{The extraction of the patch-related blocks from arbitrary binaries and the images with symbol table.}
  \label{patch_related_block}
  % \vspace{-0.15in}
\end{figure}

\vspace{0.05in}
\noindent\textbf{For arbitrary binaries.}
With the comparison between the existing diff tools, we choose DeepBinDiff~\cite{duan2020deepbindiff} to extract the patch-related blocks. 
The original DeepBinDiff directly detects basic blocks with distinct semantics by matching inter-procedural control-flow graphs (ICFG) in the program-wide code representation, which is precise but leads to high overhead.
Thus, we modify the workflow of DeepBinDiff to reduce the false positive rate and improve computation efficiency.
% For arbitrary patched and unpatched binaries, we leverage a modified version of DeepBinDiff \cite{duan2020deepbindiff} to extract patch-related assembly blocks,
As shown in Figure \ref{patch_related_block}(a), our modified DeepBindiff first narrows down the search scope to the function level and then filters discrepant basic blocks within patch-related functions.
Such an approach works for arbitrary binaries, even for striped binaries, which do not include symbol table information.

\vspace{0.05in}
\noindent\textbf{For images with the symbol table.}
Our \TN{} can also be used to identify security patches within released images.
For pre-patch and post-patch images, the workflow of patch-related block extraction is shown in Figure~\ref{patch_related_block}(b). 
As the symbol table is included in most Linux-based kernel images~\cite{zhang2018precise}, we can easily locate the functions by checking the entities with the same name.
Then, we perform syntax-based function similarity detection method to filter the identical functions.
Finally, we utilize the modified DeepBinDiff to extract patch-related blocks within functions.

\vspace{0.05in}
\noindent\textbf{For the ground truth in the Benchmark.}
We leverage the information of source code changes in commit messages and the debug information within binaries to locate patch-related blocks precisely. 
Specifically, we build a mapping between instruction and source code line number by using the DWARF information~\cite{dwarf2021debug}.
Then, we could precisely locate each patch-related block according to the source code changes.

% \vspace{-0.1in}
\subsection{Graph Extraction}

\begin{figure}[ht]
    \centering
    \includegraphics[width=2.8in]{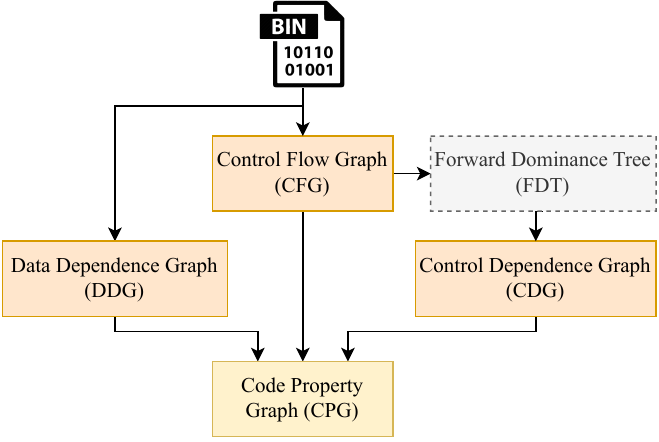}
    \caption{The code property graph (CPG) generation from the CFG, CDG, and DDG.}
    \vspace{-0.1in}
    \label{fig:cpg}
    \vspace{-0.1in}
\end{figure}

In addition to patch-related block sets, the impact of patches on program semantics is also reflected in the changes in program flow and dependency. 
In other words, a patch also contains the relationship information between basic blocks.
Code property graph (CPG) is a language-agnostic intermediate program representation, which merges multiple code graph representations into one queryable graph database~\cite{Yamaguchi2014Joern}. 
To include all possible relations, our CPG-based representation merges three types of graphs (i.e., control flow graph (CFG), control dependence graph (CDG), and data dependency graph (DDG)) into a single joint data structure.
As shown in Figure~\ref{fig:cpg}, CFG represents all the possible traversed paths during program execution, and DDG represents the data access relation between statements. 
Noting that CDG is derived from CFG via inferring the forward dominance tree (FDT), which means that node B dominates node A if node A determines whether node B is executed. 
In our paper, CFG contains the direct dominance relationship, while CDG focuses on the indirect dominance relationships to avoid overlapping connections. 

% The CPG merges three compiler representations (i.e., CFG, CDG, and DDG) into a single joint data structure. 
% AST is a code representation generated by the syntax analysis of a compiler. 
% In other words, CPG represents all the possible traversed paths during program execution. 
% CDG and DDG represents control dependency, and data dependency, respectively [xx]. 
% PDG comprises of control dependency graph (CDG) and data dependency graph (DDG) to represent the control and data dependencies, respectively [**]. 
% By containing all the information of control flow, control dependency, intra-procedural data dependency, and program syntax, CPG provides a comprehensive view for code static analysis.

% \vspace{0.05in}
% \noindent\textbf{Graph Generation}
% After extracting patch-related assembly blocks, we derive connected sub graphs to cover all patch blocks.
Given a patch-related block set, \TN{} constructs the CPG-based graph in two steps.
First, we derive a CFG to only connect all patch-related blocks, which indicates the internal connectivity of the patch.
% Such a subgraph is to establish internal connectivity.
Second, we extend a larger graph to cover all patch-related blocks and connect them to their neighborhood basic blocks. 
That is, we add more context basic blocks to the graph, as long as these blocks are connected to patch-related blocks in the CFG, DDG, or CDG.
Such a graph representation further implies the external impact of the patch. 
As shown in Phase II of Figure~\ref{fig:security_detection}, the orange node in the graph denotes the patch-related blocks and the pale yellow node denotes the context blocks.
% 这一步可以确定补丁的作用域。

% Specifically, we consider two kinds of graph features: 1) \textit{subcfg}, derives only from control flow graph (CFG) and covers all patch-related blocks; 2) \textit{slicedsubcfg}, derives from CFG, control dependence graph (CDG) and data dependence graph (DDG) and covers all patch-related blocks and other affect sliced blocks. 
% The sliced blocks exists data or control dependence with patch-related blocks.

% The \textit{slicedsubcfg} graph features are more informative, which could capture more accurate characteristics of security patches.

% The workflow of sub graph extraction is shown in Figure . This approach extracts the assembly sub graph from assembly and binary files. 
% It leverages debug parser to extract the mapping relation between sequence with instruction, and instruction with line number. Then, it uses the disassembly tool to extract the control flow graph information and builds mapping relation between sequence with instruction, instruction with block, and block with edge. Finally, based the patch information, we perform sequence matching to extract patched sub graph.
% In this section, we hold the hypothesis that the instruction sequences within assembly files are the same with those in binary files. 

% \begin{figure}[h]
%     \centering
%     \includegraphics[width=0.5\textwidth]{images/patched_sub_graph_extraction.pdf}
%     \caption{The workflow of sub graph generation.}
%     \label{fig:sub_graph_extraction}
% \end{figure}
\vspace{0.05in}
\noindent\textbf{Graph Slicing.}
We adopt the graph slicing method to extract the relevant node connections and detect all relevant basic blocks in the CPG representation.
Given a pair of pre-patch and post-patch binaries, we first extract their CFG, CDG and DDG from the assembly code.
For the internal graph, we match the patch-related blocks in the program-wide CFG to retain internal connections.
For the entire graph, we design a graph slicing algorithm.
% , as presented in the Algorithm~\ref{alg:prfunction}.
% The Algorithm~\ref{alg:prfunction} presents the workflow of the graph slicing. 
% The workflow of extracting slicing information is presented in Algorithm~\ref{alg:prfunction}. 

% we leverage disassembly tool to extract CFG, CDG and DDG.
The algorithm takes the patch-related block set ($BB$), the graph relations including CFG, CDG, and DDG ($\{G_{cfg}, G_{cdg}, G_{ddg}\}$), and the slice stride ($n$) as inputs, and outputs the sliced CPG-based representation.
Traversing each basic block from the block set, \TN{} first traces the control and data dependency relations between blocks. That is to find the predecessor and successor of the current block in CFG, CDG, and DDG. 
% \TN{} first builds the mapping from each instruction to its corresponding block $instToBlock$.
% Then, the control and data dependency relations between blocks are built according to CDG and DDG (Line 3-13). 
% The $blockToControlPreds$ and $blockToControlSuccs$ represent the relations from the block to its predecessor and successor blocks in the CDG, respectively.
% Similarly, the $blockToDataPreds$ and $blockToDataSuccs$ represent the predecessor and successor relations of the current block in DDG.
Correspondingly, We will create two collections ($forSlices$ and $backSlices$) to store the predecessors and successors of all blocks, respectively.
% the $forSlices$ contains all successor nodes and the $backSlices$ contains all predecessor nodes.
Next, \TN{} iteratively performs both forward and backward slicing to extract all neighborhood blocks.
The slice stride $n$ is a configurable parameter that represents the maximum distance from a context node to the nearest patch-related node.
% a patch-related node to an external node.
The slicing process will terminate after $n$ iterations or until the sliced graph structure converges.
Finally, we connect the patch-related blocks $BB$, the forward sliced blocks $forSlices$, and the backward sliced blocks $backSlices$ as the final CPG-based representation $\boldsymbol{G}$.

\subsection{Embedding Generation}
In the graph representation, the extracted nodes and edges cannot be directly fed into the deep learning model. 
Thus, we convert the graph topology into an adjacency matrix, which can represent the node connection information.
For the edges and nodes, we embed their attributes into the numeric matrices, which are represented as node embedding $\boldsymbol{N}$ and edge embedding $\boldsymbol{E}$.
% Graph embedding is a type of graph representation where all the nodes, edges, and their features are transformed into a unified vector. 
Given a CPG representation $\boldsymbol{G}=\{\boldsymbol{E},\boldsymbol{N}\}$, 
we first define an adjacency matrix $\boldsymbol{A}$ to denote the connectivity of $\boldsymbol{G}$, where $\boldsymbol{A}_{ij} = 1$ represents that node $i$ and node $j$ is connected, else $\boldsymbol{A}_{ij} = 0$ if they are unconnected. 
Because there are three subgraphs in the CPG, i.e., CFG, CDG, and DDG, $\boldsymbol{A}^{(k)}$ denotes the connectivity in the \textit{k}-th subgraph.
Note that, each subgraph in \TN{} is a directed graph.
For the edge representation, we denote the matrix of edge embeddings as $\boldsymbol{E}$, where $\boldsymbol{E}_d^{(k)}$ is the \textit{d}-th edge in the \textit{k}-th graph. 
$\boldsymbol{E}^{(k)}$ is a vector that contains all edges of the \textit{k}-th subgraph.
% Because $E_d^{(k)}$ can be either 0 or 1, we generate a masked adjacency matrix $M^{(k)}$ according to $E^{(k)}$, where $M^{(k)}_{ij} = 1$ if the edge connecting node $i$ and node $j$ has the \textit{k}-th attribute of 1, else $M^{(k)}_{ij} = 0$. 
% $M^{(k)}$ can be used to reflect the node connections of the \textit{k}-th subgraph.
For the node representation, we denote the matrix of node embeddings as $\boldsymbol{N}$, where $\boldsymbol{N}_i$ is the \textit{i}-th basic block in the CPG. Also, nodes can be shared by different subgraphs. 
For each node, we represent them with semantic embedding vectors.

% \clearpage
\vspace{0.05in}
\noindent{\large{\textbf{\textit{A. Node Embedding}}}}
\vspace{0.03in}

\noindent As shown in Phase III of Figure~\ref{fig:security_detection}, we convert all nodes into the numeric representation, including the patch-related blocks and the context blocks.
In binary code analysis, the most common numeric representation is the statistic-based features, such as counting distinct tokens in basic blocks. These tokens refer to operators and operands in instructions~\cite{alon2019code2vec}.
However, such an embedding method focuses more on the syntactic features but neglects the semantic information, which may be not fine enough for the subtle changes in security patches.
Unlike source code, binary code is also subject to compiler optimizations. 
The same source code may be compiled into different binaries with different optimization configurations.
Therefore, the LM-based embedding model is utilized in \TN{} due to its powerful semantic understanding capability~\cite{radford2018improving,devlin2018bert,lewis2019bart,feng2020codebert,he2022binprov}.

\vspace{0.08in}
\noindent\textbf{Embedding Model.}
% describe the BERT-based embedding model
The \TN{} employs a BERT-based embedding model~\cite{pei2020xda} to learn the instruction semantics in the basic blocks.
The BERT model can learn the context semantics of the instructions through specially designed self-supervised training tasks~\cite{li2021palmtree,pretrained_model}.

\begin{figure}[t]
    \centering
    \includegraphics[width=3.3in]{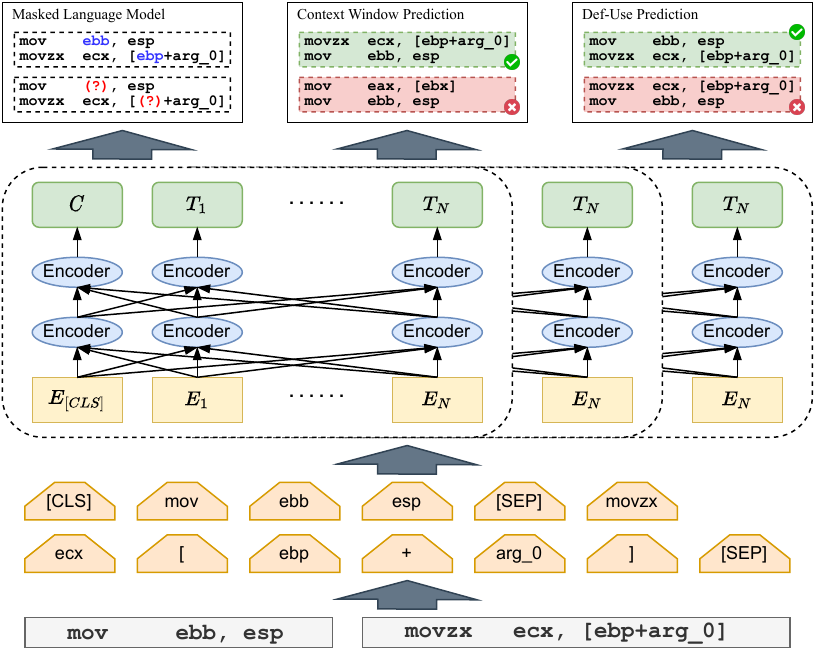}
    \caption{The pretraining of BERT-based embedding model.}
    \label{fig:embed}
\end{figure}

Exploiting the transformer framework and attention mechanism, BERT can learn long-range contextual semantics of the instruction sequences in basic blocks.
From the perspective of model structure, BERT is stacked by multiple transformer encoder units that share the same architecture.
The BERT model consists of 12 transformer layers, which share the same architecture.
As shown in Figure~\ref{fig:embed}, the BERT tokenizes the input instruction sequences and encodes them as initial embeddings.
The model will continue updating the embeddings via the transformer encoders. 
To distill the learning capability of the model on the instruction semantics, we need to train the model via well-designed training tasks.
We denote the \textit{j}-th layer embedding as ${EM}_j$ and the encoder unit as $f_e$, then the output embedding in each encoder layer is expressed as ${EM}_{j+1}=f_e({EM}_{j})$.

% The updated embedding is known as contextualized embedding [xxx].
%\cite{pennington2014glove,devlin2018bert,mikolov2013distributed,mikolov2013efficient,peters2018deep}.
%
% So the output of the BERT model can be expressed as $E_{out} =f_e(E(b))$, where $f_e$ represents the whole embedding model.
% The transformers in BERT share the same architecture, so the output of the BERT model can be expressed as $E_{final} =f_e(E(b))$, where $f_e$ represents the whole embedding model, which consists of 12 transformers.
% $E_{out} \in \mathbb{R}^{N\times M}$, where $N$ denotes the length of the input sequence and $M$ denotes the dimension of the token embedding.

% For the semantics of the instruction, we leverage the LM-based embedding model to learn and represent the context semantics and syntax directly from assembly instructions. Existing work shows that the language model can capture subtle semantic changes in binary code. We, therefore, generate LM-based embeddings as attributes of nodes in the graph representation, where each node represents a basic block related to the patch.

\vspace{0.08in}
\noindent\textbf{Tokenization.}
% Tokenization是生成embedding的第一步，基本方法是将一条指令分解为one opcode and several operands. 然而， operand可以进一步细分。例如，
Before we feed the instruction sequences into the embedding model, we need to tokenize them.
% Tokenization is the first step to generating embedding. 
The basic way is to decompose an instruction into one opcode and multiple operands. However, the operand can be further subdivided.
For example, given an instruction "\texttt{mov rax, qword [rsp+0x58]}", we divide it into "\texttt{mov}", "\texttt{rax}", "\texttt{qword [rsp+0x58]}" originally. 
% 但是操作数"qword [rsp+0x58]"过于复杂以至于可能会导致OOV (Out-Of-Vocabulary) problem。所以我们将操作数细分为所有可能的元素，包括：寄存器，立即数，保留字，和运算符
Note that the operand "\texttt{qword [rsp+0x58]}" is too complicated to cause an OOV (Out-Of-Vocabulary) problem. 
Thus, we subdivide operands into more basic tokens including registers (e.g., \texttt{rax} and \texttt{rsp}), constant (e.g., \texttt{0x58}), reserved words (e.g., \texttt{qword}), and operators (e.g., \texttt{+}, \texttt{[}, and \texttt{]}).
% In other words, we consider each instruction as a sentence and decompose the operands into more basic elements.
% As introduced earlier, unlike Asm2Vec which splits an instruction into opcode and up to two operands, we apply a more fine-grained strategy. For instance, given an instruction ``mov rax, qword [rsp+0x58]'', we divide it into “mov”, “rax”, “qword”, “[”, “rsp”, “+”, “0x58”, and “]”. In other words, we consider each instruction as a sentence and decompose the operands into more basic elements.

\vspace{0.08in}
\noindent\textbf{Input Embedding.}
The input representation of the BERT model is a composite embedding, concatenated by three parts: token sequence, segment sequence, and position sequence. 
The token sequence represents the tokenized instruction sequence. 
The segment sequence is defined to locate which instruction each token belongs to, while the position sequence denotes an integer sequence encoding the position of each token in the instruction. 
% The segment and position flags can effectively assist BERT in understanding the belonging and positional relations of each token. 
% we use one-hot encoding to generate byte embedding vectors for these 3 input sequences, and concatenate them into a unified input vector formulated as: $E = {f}_{concat}(E_b, E_p, E_s)$. 

\vspace{0.05in}
\noindent\textbf{Pre-training task.}
The BERT model proposed two representative pre-training tasks, that is, masked language model (MLM) and next sentence prediction (NSP).
% MLM can help BERT learn bidirectional contextual semantics, while NSP boost the capability of BERT in sentence handling and paragraph matching.
MLM will train the model to predict masked tokens in the input sequence, while NSP will train the model to predict the next sentence according to the former one. 
Liu~\etal~\cite{Liu2019RoBERTaAR} found that the BERT model pre-trained only with MLM can outperform that pre-trained with both tasks. 
Therefore, we adopt the MLM task to train the BERT model.

In addition to the MLM task, there are also dedicated training tasks proposed for learning assembly code semantics: Context window prediction (CWP)~\cite{li2021palmtree} and Def-use prediction (DUP)~\cite{Lan2020ALBERT}. 
As shown in top boxes of Figure~\ref{fig:embed}, the CWP task is to predict if two given instructions co-occur within a context window, where the window size is preset. 
% and the width of the window is preset.  
The DUP task is to determine the relative order of two given instructions.
The CWP can capture the control flow information between instructions, while the DUP can grasp the data dependency information across instructions.
In this paper, we choose the MLM, CWP, and DUP tasks to train our BERT-based embedding model and then generate the node embedding.

\vspace{0.05in}
\noindent{\large\textbf{\textit{B. Edge Embedding}}}
\vspace{0.03in}

\noindent Edge embedding is used to reflect the connectivity between two nodes. The edges in the graph representation involve three types of relationships, i.e., the control flow, control dependency, and data dependency.
% , i.e., version information and edge types. The version information refers to whether the edge is present only in the pre-/post-patch version or in both versions. 
The edge type refers to which type of connections it belongs to.
For example, $[0,0,1]$ means the edge represents the DDG, rather than CFG or CDG.
Note that there may be multiple edges (e.g., one from CDG and one from DDG) between two nodes. 
% Apart from these three types of connections, there are extra dimensions to indicate if the edge is present in the pre-patch version and the post-patch version, respectively. 
Therefore, the edge embedding is designed as a 3-dimensional binary vector. 

% If the edge belongs to both versions, the first two bits will be (1, 1). 
% The last three bits indicate if there are any CDG, DDG, or AST edge between the current two nodes, respectively.

% \vspace{0.05in}
% \noindent\textbf{Vectorization of Graph Representation}

\subsection{Graph Representation Learning}
After converting the graph representation into the embedding format, \TN{} will feed the embedding matrices into a detection model, which can provide a powerful capability of graph understanding by the graph neural networks.

% \vspace{0.05in}
% \noindent\textbf{Graph Model.}  
% General Design
% PatchGNN is based on graph learning, which provides a great capability of graph classification by neural networks. 
As shown in Phase IV of Figure~\ref{fig:security_detection}, our graph model adopts a siamese structure consisting of two graph convolution networks. 
Such a structure is conducive to processing pre-patch and post-patch graph representations separately, which can emphasize the differences caused by patches while maintaining similar contexts before and after patches.
Each graph network is stacked with the convolution layers and the pooling layers.

\begin{figure}[t]
    \centering
    \includegraphics[width=3.3in]{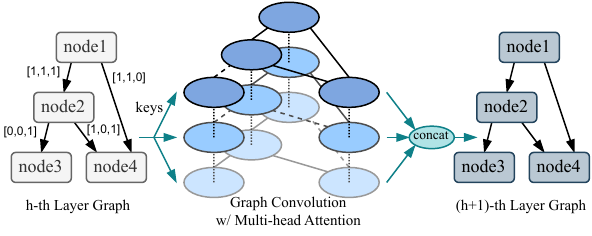}
    \caption{Graph convolutional layer w/ multi-head attention.}
    \label{fig:gcn}
\end{figure}

As displayed in Figure~\ref{fig:gcn}, the graph convolution layer is a variant of the normal convolution layer in the CNN networks, updating the node embeddings by the message propagation among the neighboring nodes.
Note that, we limit the number of convolution layers to three because more convolution will cause the graph over-smoothing issue~\cite{chen2020measuring}, which means excessive convolution may smooth the node attributes in the graph so that their information is permanently lost.
% to become so consistent that they cannot be distinguished.
% 过度的卷积会导致图中节点的属性趋于一致以至于无法区分。
Behind the 3-layer graph convolution, we use a graph pooling layer to aggregate all attributes to get the graph embedding.
% graph embeddings can be obtained by graph pooling and vector concatenation. 
% GCNs perform similar operations where the model learns the features by inspecting neighboring nodes. 
% The graph convolutional layers will update the node embeddings of PatchCPG with the neighborhood information in different subgraphs. 
% We only use 3 convolutional layers in PatchGNN because more convolution will lead to graph over-smoothing. 
Finally, a binary classifier constructed by multiple-layer perceptron (MLP) is utilized to convert the graph embeddings into predicted labels. 
% Next, we introduce the mechanism of graph convolution and classifier respectively.

% Multi-attributed Graph Convolution
\vspace{0.05in}
\noindent\textbf{Multi-Attributed Graph Convolution.}
The CPG is a composite graph with diverse edge attributes, representing different connections in CFG, CDG, and DDG.
Therefore, we adopt the multi-attributed graph convolution mechanism to process multiple edge types. 
That is, the graph convolution layer has three channels to process the edge embeddings from different subgraphs.

When the convolution layer updates the node embedding, it collects the embeddings from neighborhood nodes with different subgraph connections.
In other words, the graph connection in each convolutional channel depends on the graph type, respectively.
% one node may be updated up to 3 times in one convolution layer.
% it needs to divide different subgraphs according to the type of connections. The neighbor nodes can only be selected from the same subgraph.
Therefore, we train different weights for the each channel of subgraph to learn semantics from different connection relations.
% 因此，我们实际上为每一种子图的图学习训练了不同的权重，从而可以表达不同的连接关系。
% In other words, each edge embedding dimension will be fed into different channels of the convolution layer.
% our graph representation is a heterogeneous graph. To better leverage the hidden knowledge, we construct the PatchGNN model with a multi-attributed graph convolution mechanism. 
% views each dimension of edge embeddings as an individual convolution channel.
% In this paper, each dimension in the edge embedding represents the connection in CFG, CDG, and DDG. 
% Therefore, 在卷积层从邻居节点获取信息更新当前节点的embedding之前，需要划分不同的子图以选择邻居节点来自同类型的连接。
% For each node in PatchCPG, the convolutional layers in PatchGNN tend to gather information from its neighbors via different types of edges. 
% Due to the different roles of edge types, we cannot use a set of unified weights to learn the detection model. Each dimension of the edge embeddings can indicate different relationships in PatchCPG, i.e., pre-patch/post-patch connections, control/data dependencies, and AST graph. 
% As shown in Figure 5, each edge can have multiple attributes while each type of attributes can be used to construct a subgraph. As the information in each dimension corresponds to a subgraph, we can view each dimension as an individual convolution channel. 
% In the convolutional layers of PatchGNN, we process the structural information of each channel individually and aggregate the processed information of all subgraphs.
Formally, the convolution update can be formalized as follows:
\begin{equation}
    \boldsymbol{N}^{(h+1)}= \Vert_{k=1}^{K} \boldsymbol{\sigma}\left(\left(\boldsymbol{A}^{(k)}+\boldsymbol{I}\right) \cdot \boldsymbol{N}^{(h)} \cdot \boldsymbol{W}_{h}^{(k)}\right),
\end{equation}

\noindent where $\boldsymbol{A}^{(k)}$ is the adjacency matrix of the \textit{k}-th subgraph, and $I$ is an identity matrix with the same size of $\boldsymbol{A}^{(k)}$. 
$\boldsymbol{N}^{(h)}$ denotes the node embedding in the \textit{h}-th convolution layer.
% where $\boldsymbol{A}$ is the adjacency matrix of the CPG-basd graph, $I_N$ is the identity matrix of size $N$ , and $\odot$ is the Hadamard product.
% $\boldsymbol{X}^{(h)}$ is the node embeddings in the h-th convolution layer.
$\boldsymbol{W}^{(k)}_h$ is the convolution weights of the \textit{k}-th subgraph in the \textit{h}-th layer, which will be trained via the backpropagation. 
$K$ is the total number of subgraphs and $\boldsymbol{\sigma}$ is the activation function.
Finally, the updated node embeddings from different subgraphs will be aggregated to the convolution result $\boldsymbol{X}^{(h+1)}$ by the vector concatenation. 
The adjacency matrix $\boldsymbol{A}^{(k)}$ can be thought of as a filter that provides individual attention for each subgraph to train convolution parameters.

% Security Patch Classifier
\vspace{0.05in}
\noindent\textbf{Security Patch Classifier.}
Behind the convolution layer, a pooling layer is used to reduce the embedding dimension.
Then, the graph embeddings from both GCN branches are concatenated as the input embedding for the final classifier.
Also, a dropout layer is performed as a regularization method to prevent over-fitting in the model training. 
% Since the output of convolutional layers is graphs, we need further processing to obtain the final predictions. 
% First, the graph pooling layers are leveraged to reduce the data dimension and acquire the graph embeddings. 
% In our design, we use both mean pooling and max pooling to obtain two graph representations, each of which is a high-dimensional vector. 
% Then, these two graph representations are concatenated to construct a final graph embedding, which contains all information on the nodes, edges, and their features in a PatchCPG. 
% Afterward, a dropout layer is performed as a regularization method to prevent over-fitting in the model training. 
To determine if a patch is security-related, a 3-layer fully connected network is built to transform the graph embedding into a binary probability output ($p_0, p_1$), where $p_0 + p_1 = 1$. The higher probability in the output indicates that a patch instance falls into the category of security/non-security patch.

%% file: 04.implementation.tex
\section{Implementation}
In this section, we first introduce how to build the binary patch dataset from a set of pre-collected source code patches.
Then, we discuss the implementation of four phases in \TN{}, including patch data pre-processing, graph extraction, embedding generation, and graph representation learning. 
% We have implemented a prototype of \TN{} via Python scripts.
% The dataset generation component contains 710 line Python code.
% The sub graph extraction component is built on the top of Angr~\cite{angr2021dpython} framework and contains 850 line Python code.

\subsection{Building Binary Patch Dataset}
\label{Sec: dataset}
% \subsection{Dataset Generation}
We build the binary patch dataset based on a pre-collected source code patch dataset named PatchDB~\cite{wang2021PatchDB}.
We obtain the patch information from the PatchDB and download the corresponding source code files from the GitHub repositories.
In this paper, the most challenging part is to compile all patch-related source code files from different software repositories, which require different compile commands and various suitable building environments.
We manually prepare the building environments for different software in isolated Docker containers to avoid dependence conflicts, which can provide the build logs.
Then, we use a script to automatically extract patch-related source code files from GitHub repositories, and select the correct compile commands and dependent files from the build logs.
For versatile evaluation of \TN{} on this binary dataset, we leverage two well-known compilers (i.e., Clang and GCC) to compile these programs into binaries with different optimization levels (i.e., \texttt{O0}, \texttt{O1}, \texttt{O2}, \texttt{O3}, and \texttt{Os}), respectively.
Furthermore, we generate the LLVM IR and the assembly code for each patch to facilitate subsequent analysis work.

% \begin{figure}[h]
%     \centering
%     \includegraphics[width=3.33in]{images/commands.pdf}
%     \caption{The compile command for a Linux kernel case. The italic font represents the dependent files, while the grey font donates the compiler settings.}
%     \label{fig:compiler_command_linux}
% \end{figure}

\vspace{0.05in}
\noindent\textbf{Compilation Command Database Preparation.}
% To obtain binary patterns of security patches, we need to map the patched source code into assembly code, which requires a process of first compiling and then disassembling.
% Assuming that we are given one source code C file, we then need to find all dependent files (.h) and correct compiler settings.
Given a source code file for a C program, we need to compile the program based on two types of information, i.e., the dependent files and compiler configuration.
The dependent files include the header files and linked source files, while the compiler configuration refers to the options in the compile command line.
%
% Figure~\ref{fig:compiler_command_linux} shows an example of the compile command for the patch-related code files (i.e., \texttt{ mm/pagewalk.c}) of CVE-2017-16994. 
% The dependent files and file paths are specified via the options "\texttt{-isystem}", "\texttt{-I}", and "\texttt{-include}". 
% The command option starting with "\texttt{-W}" usually controls the warning and error message, and those starting with "\texttt{-f}" refer to optimization.
It is not trivial to seek such complicated information automatically, due to three main challenges.
First, the dependent files usually exist at different locations (e.g., \textit{./deps/lua/src}, \textit{./src/lib/openjp2}, \textit{./eglib/src}); thus, complex path relations are difficult to untangle in a general way.
Second, some dependent files are generated during the building process, rather than exist before compilation (e.g., \textit{config.h}, \textit{kdb\_ldap.h}, \textit{asm/linkage.h}).
Third, the format of configuration scripts for automated compilation tools can be largely different (e.g., \textit{autoreconf}, \textit{autogen.sh}).
% When performing the task of automatically building software, we enter another bigger challenge. 
% That is, software usually relies on specific build configurations (e.g., \textit{autoreconf}, \textit{autogen.sh}) and requires various different preliminary basic libraries. 

% We propose one generic approach for seeking dependent files. 
% We empirically collect necessary tool list and configuration command list for common softwares.
% When building software repositories, we first install these necessary tools. 
% Then, the prepared configuration commands are executed one-by-one. In this process, the invalid configuration command will not influence the build process. 
% Finally, the \texttt{make} command is executed to build the whole software.
% For seeking dependent files, we parse ``Makefile'' to extract dependent files, add all sub directories in the frequently-used directory including ``include'', ``src'', ``lib'', ``build'', and manually cover many specific files.
To solve the above challenges, we propose an automatic extraction system to extract complete compile commands from the build logs.
We first manually build the software to collect the build logs. 
Note that different versions of the same software may also have different dependencies and compilation configurations.
Thus, we build isolated Docker containers for each version to avoid dependence conflicts.
The build log is parsed into a queryable JSON format.
For each software, we build a compile command database to manage the compilation information for each source code file. 
%The construction of compiler command database is shown in Figure \ref{fig:compiler_command}.

% Then, a log parser module is leveraged to build the compiler command database from the build log. 
% The database contains compiler commands for each source code file.
% The compiler command consists of dependent header files and file paths (e.g., \texttt{-I}, \texttt{-D}) and compiler options (e.g., \texttt{-fsanitize=kernel-address}), as shown in Figure \ref{fig:compiler_command_linux}.

% Considering the huge size, the compilation of the Linux kernel is much more complex.
% % Figure~\ref{fig:compiler_command_linux} show an example of the compiler command for CVE-2017-16994 (in kernel version v4.15). 
% The patched source code file for this vulnerability is \textit{mm/pagewalk.c}. 
% From Figure~\ref{fig:compiler_command_linux}, we observe that compiler settings and dependent files are too diverse to be derived automatically. 
% The compiler option \texttt{-nostdinc} means that do not search the standard system directories and compiler built-in directories for dependent files. 
% The compiler option \texttt{-fsanitize*} represents the check for various forms of undefined or suspicious behavior within the source code.
% The option \texttt{-no-integrated-as } represents disabling the use of the integrated assembler.

% \begin{figure}[h]
%     \centering
%     \includegraphics[width=0.4\textwidth]{images/compiler_command_database.pdf}
%     \caption{The construction of compiler command database.}
%     \label{fig:compiler_command}
% \end{figure}

\vspace{0.05in}
\noindent\textbf{Compilation Target.}
With different compilation configurations, the same source code can even generate diverse assembly code, as demonstrated in Section~\ref{Sec:binaryCode}. 
To evaluate if \TN{} can handle the changes on compilation configurations, we leverage Clang~\cite{clang} and GCC~\cite{GCC} with different optimization levels to compile the source code.
% the Linux kernel from different versions (v3.0-v5.8), respectively. 
However, using different compilers and optimization levels means changing the original build environments, which may cause compilation failure.
For instance, the earlier versions of Linux kernel can not be compiled with Clang~\cite{wang2018check}, due to the incompatibility with the LLVM toolchain.
To solve this issue, we choose the \texttt{allyesconfig}~\cite{pagani2021autoprofile} to cover as many source code files as possible. 
The basic idea is to compile the Linux source code module by module, while a module will be skipped if it cannot be compiled successfully. 
The \texttt{allyesconfig} totally includes 16,599 modules, among which we successfully generate LLVM IR for 16,514 modules. 
% Furthermore, we use Clang with the command option "\texttt{-g -fno-integrated-as -fno-inline}", where the "\texttt{-fno-integrated-as}" and "\texttt{-fno-inline}" options are used to disable integrated and inline assemblers. 
Furthermore, the "\texttt{-g}" option is used to generate debug information, which is useful to investigate the code lines and to further locate the patched assembly code snippets.

\begin{figure}[t]
    \centering
    \includegraphics[width=3.1in]{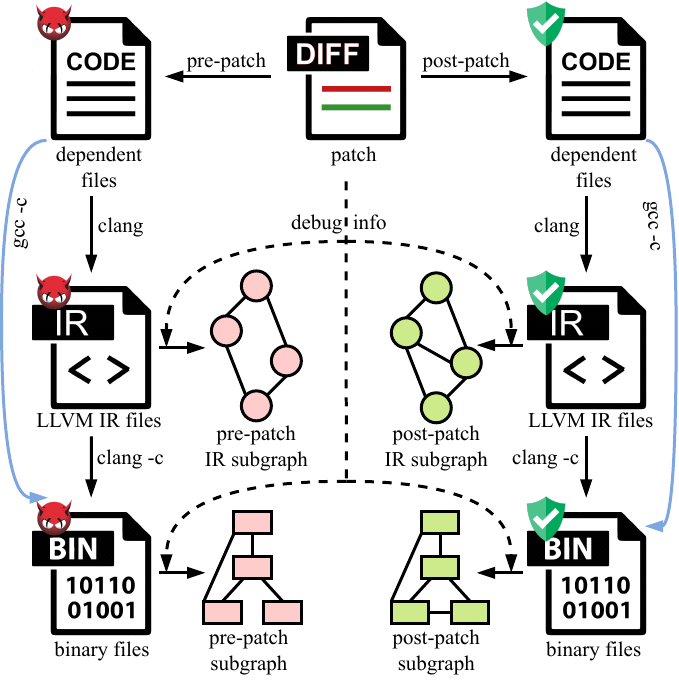}
    \caption{The workflow of binary patch generation via Clang or GCC.}
    \label{fig:automatic_approach}
    \vspace{-0.1in}
\end{figure}

\vspace{0.05in}
\noindent\textbf{Automatic Binary Patch Generation.}
The workflow of binary patch generation is shown in Figure~\ref{fig:automatic_approach}. 
Given the patch information, we can locate the pre-patch and post-patch software versions and obtain patch-related source code files that contain the changed code lines.
For the sake of efficiency, we only compile the patch-related source code files.
According to the version information and the source code filenames, we can query the corresponding compile commands from the compilation command database for the pre-patch and post-patch source code files.

Clang is able to translate the source code into the LLVM IR code and directly perform optimization and program analysis according to the modular LLVM toolchain.
Therefore, in this work, we use Clang to first generate the IR code and then compile it into the binary code.
However, different from the structure of Clang, the front-end, middle-end, and back-end of GCC are coupled with each other.
GCC cannot directly perform optimization on the IR code. % like Clang.
Thus, we use GCC to compile the source code directly into the binary code.

To ensure that the binary code compiled by GCC can be converted into the same assembly and IR format as that compiled by Clang, we also use the LLVM toolchain to disassemble the binary compiled by GCC and lift it into LLVM IR code.
As demonstrated in Section~\ref{Sec:blockextract}, we locate the patch-related basic blocks in IR and assembly code using the static analysis passes provided by LLVM~\cite{altinay2020binrec, wu2020precisely}.
Also, we provide the CFG and the CPG-based graph features, which are useful in not only security patch detection but also other patch analysis tasks, such as vulnerability detection, patch presence, and hot patch generation.

\vspace{0.05in}
\noindent\textbf{Composition of Dataset.}
We retrieve target source code programs from Linux kernel and select 1,278 security patches (fixing CVE vulnerabilities) and 1,620 non-security patches labeled by the PatchDB~\cite{wang2021PatchDB}. 
PatchDB obtains the security/non-security patch information from NVD and GitHub repositories.
For this dataset, we focus on the single-purpose patches whose commit message only describes one issue, e.g., fixing a vulnerability or adding a new feature. %because these patches have a single purpose
All these patches cover hundreds of different Linux kernel versions.
For the sake of efficiency, we only collect build logs for the major version of the Linux kernel (e.g., v3.0.0 and v5.8.0).  
The building environments of the minor versions only change slightly compared with the neighboring major version, thus we can use the same building environment for consecutive minor versions (e.g., v5.6-rc1 and v5.6-rc2).

For each patch, we compile the pre-patch and post-patch versions with GCC and Clang under five different optimization levels (i.e., \texttt{O0}, \texttt{O1}, \texttt{O2}, \texttt{O3}, and \texttt{Os}). 
Because a patch may change code across multiple functions, a patch may be related to multiple graphs. The statistic information of generated patches is shown in Table~\ref{tab:dataset}.
Note that, the division ratio between the training set and the test set is 8:2 in the subsequent experiments.
% 在后续的实验过程中，训练集和测试集的划分比例是8:2。
% First, different Linux kernel versions require different software environments and generate different header files. 
% Second, building a Linux kernel for each security patch is time-consuming. 
% We build each Linux kernel version in a Docker container and save the isolated built kernel environment within Docker images.
% After making preparation for the compiler command database and the built container environment, we can automatically generate binary patches.
% All patch information was pre-collected in a source-code level 

% After filtering, we finally obtain 1278 security patches in the Linux kernel.
% For the non-security patches, we collect commits from Linux kernel repositories, manually verify 1620 patches that fix functionality issues and add new features as non-security patches.

\begin{table}[t]
\begin{center}
\caption{The statistic of the dataset used to evaluate \TN{}.}
% \vspace{-0.05in}
\label{tab:dataset}
\resizebox{\linewidth}{!}{
\begin{tabular}{c|c|c|cc|cc}
\toprule
\multirow{2}{*}{\bf Patch Type} & \multirow{2}{*}{\bf  \#Patches} & \multirow{2}{*}{\bf \#Funcs} & \multicolumn{2}{c|}{\bf \#Nodes} & \multicolumn{2}{c}{\bf \#Edges}     \\ 
\cline{4-7} 
{} & {} & {} & \multicolumn{1}{c|}{\bf Avg} & {\bf Max} & \multicolumn{1}{c|}{\bf Avg} & {\bf Max}   \\ 
\midrule
security & 1,278 & 10,630 & \multicolumn{1}{c|}{22.81}  & 840  & \multicolumn{1}{c|}{107.7}  & 7,012  \\ 
\midrule
non-security & 1,620 & 14,410 & \multicolumn{1}{c|}{30.38}  & 1,588 & \multicolumn{1}{c|}{150.19} & 16,933 \\ 
\bottomrule
\end{tabular}
}
\end{center}
\vspace{-0.05in}
\end{table}

\subsection{System Implementation}
\vspace{0.05in}
\noindent\textbf{Patched Block Extraction.}
For the training data in our self-built benchmark, we precisely locate patch-related assembly blocks via mapping relations within debug information.
For the testing data, we utilize the modified DeepBinDiff to collect patch-related blocks.
We first aggregate basic blocks into functions and then filter similar functions according to the cosine distance of function embeddings.
Next, we filter similar basic blocks in the remaining functions, which are identified patch-related blocks. 
To evaluate \TN{}, we totally extract 1,332,031 basic blocks from 2,898 patches, whose statistical information is displayed in Table~\ref{tab:dataset}.
% Then, for function pairs existing differences, we perform embeddings-based block filtering to extract patch-related blocks.
% For Linux kernel based images (e.g., Android devices), we leverage symbol table to perform function matching and then perform syntax based function filtering to filter identical functions.
% Finally, we directly leverage DeepBinDiff to perform block embeddings to extract patch-related blocks within different functions.

\vspace{0.05in}
\noindent\textbf{Graph Extraction.}
We leverage the \emph{angr} framework~\cite{angr2021dpython} to implement the graph slicing algorithm.
To extract the internal graph representation, we first build the CFG to connect all patch-related nodes.
Then, we search all relevant external nodes among all three subgraphs, starting from the patch-related nodes. 
To reduce the graph size and exclude relatively irrelevant nodes, we empirically set the node searching within 2 hops.
Also, we set the time limit as 15 minutes for each graph generation to prevent path explosion.
Finally, all subgraphs are merged into a CPG.
In total, we generate 6,457,351 edges in total for both security and non-security patches.
% For the extraction of \textit{subcfg}, we traverse the CFG to extract sub graph which connects all patch-related blocks.
% In this paper, the extraction of \textit{slicedsubcfg} mainly depend on the generation of CFG, CDG and DDG.
% During graph generation, we choose to generate three graphs only for patched functions to improve efficiency and save system resource.
% Specifically, the nodes of DDG are represented via assembly instructions, which is different from CFG and CDG. 
% Thus, after slicing extraction, we obtain two heterogeneous graphs.
% To avoid handling complex heterogeneous graphs \cite{wang2019heterogeneous}, we map the data dependence relationship of instructions into their corresponding blocks' to unify the control flow, control dependence and data dependence information of patch at same block level, which could transfer the heterogeneous graphs into a homogeneous graph with different edge attribute and simplify graph learning to capture critical pattern.
% \vspace{0.05in}
% \noindent\textbf{Parameter setting.} 

\begin{figure}[t]
    \centering
    \subfloat[O0]{
        \includegraphics[width=1.6in]{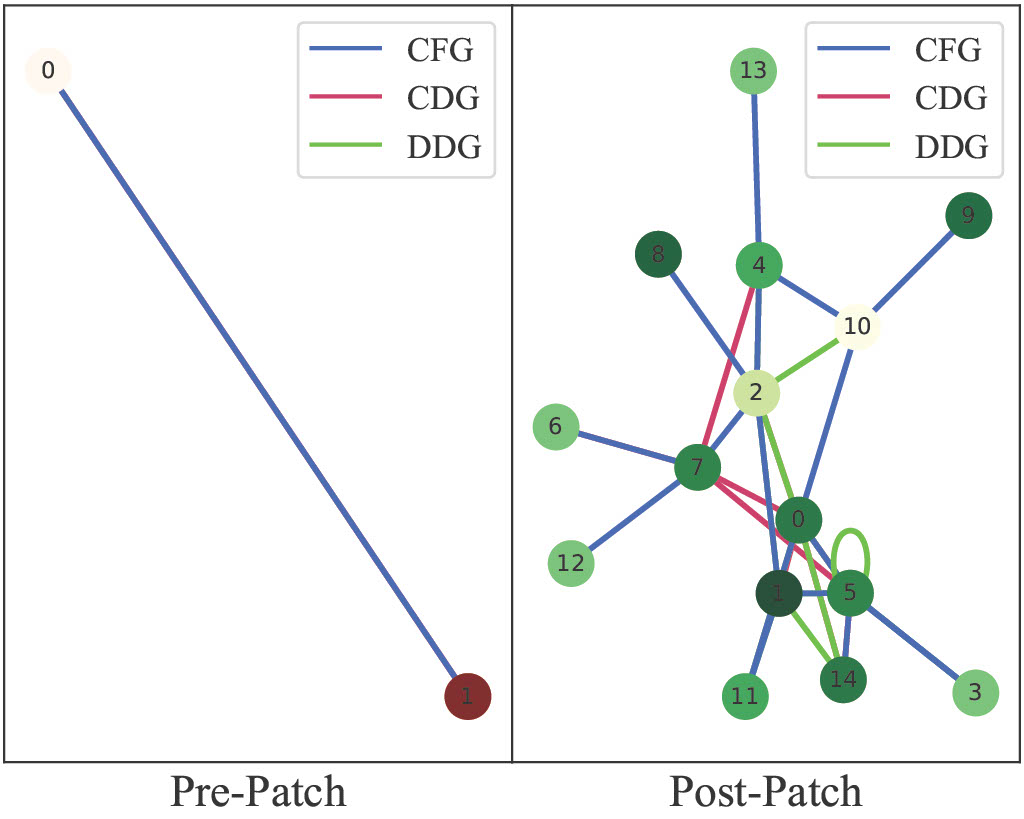}
        }% 
    \subfloat[O1]{
        \includegraphics[width=1.6in]{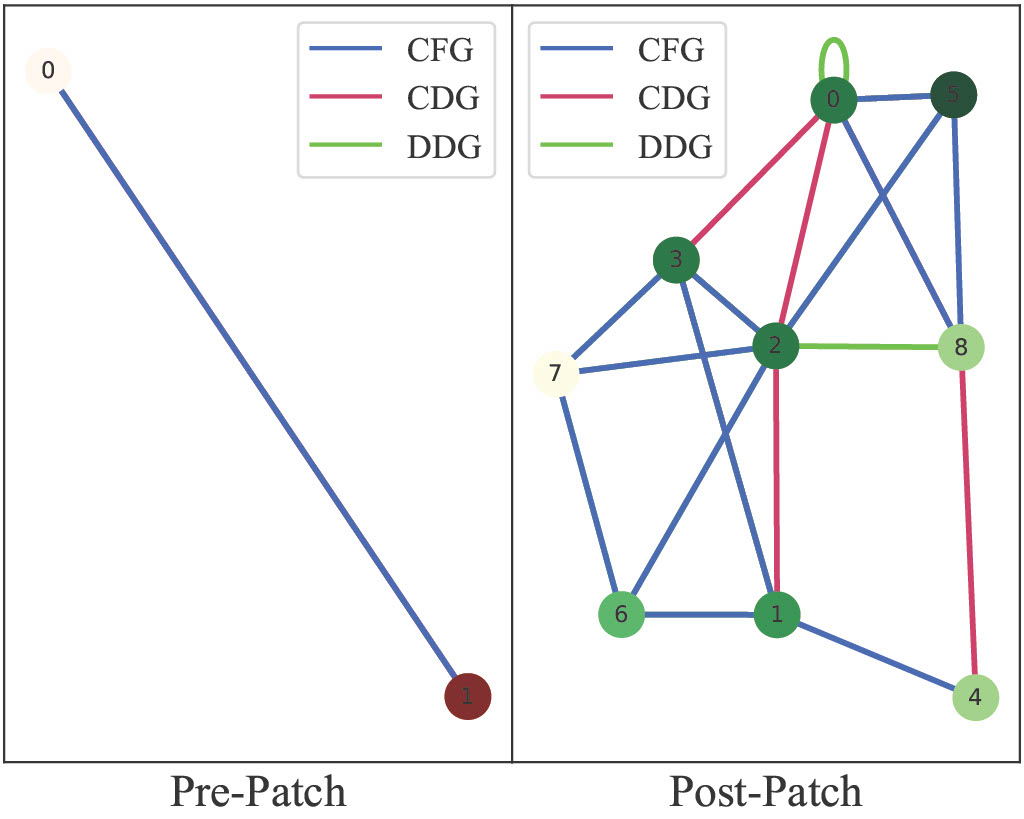}
        }% 
    \hfill
    \vspace{-0.1in}
    \subfloat[O2 (O3 is the same in this case)]{
        \includegraphics[width=1.6in]{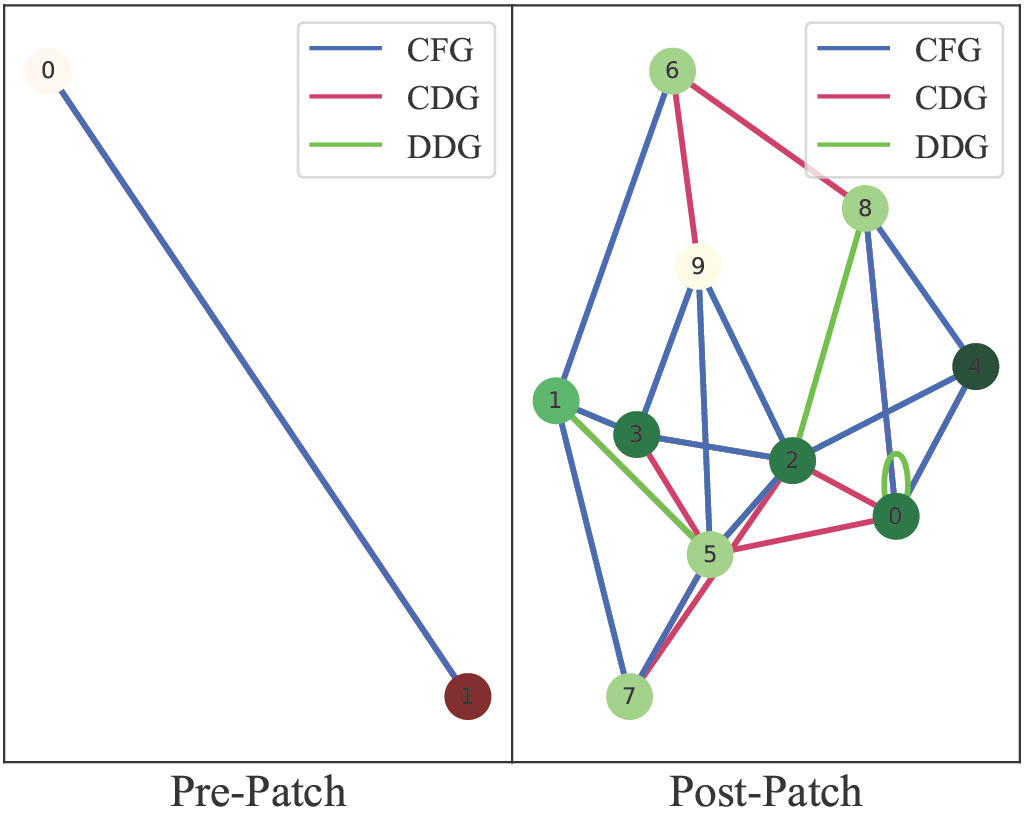}
        }% 
    \subfloat[Os]{
        \includegraphics[width=1.6in]{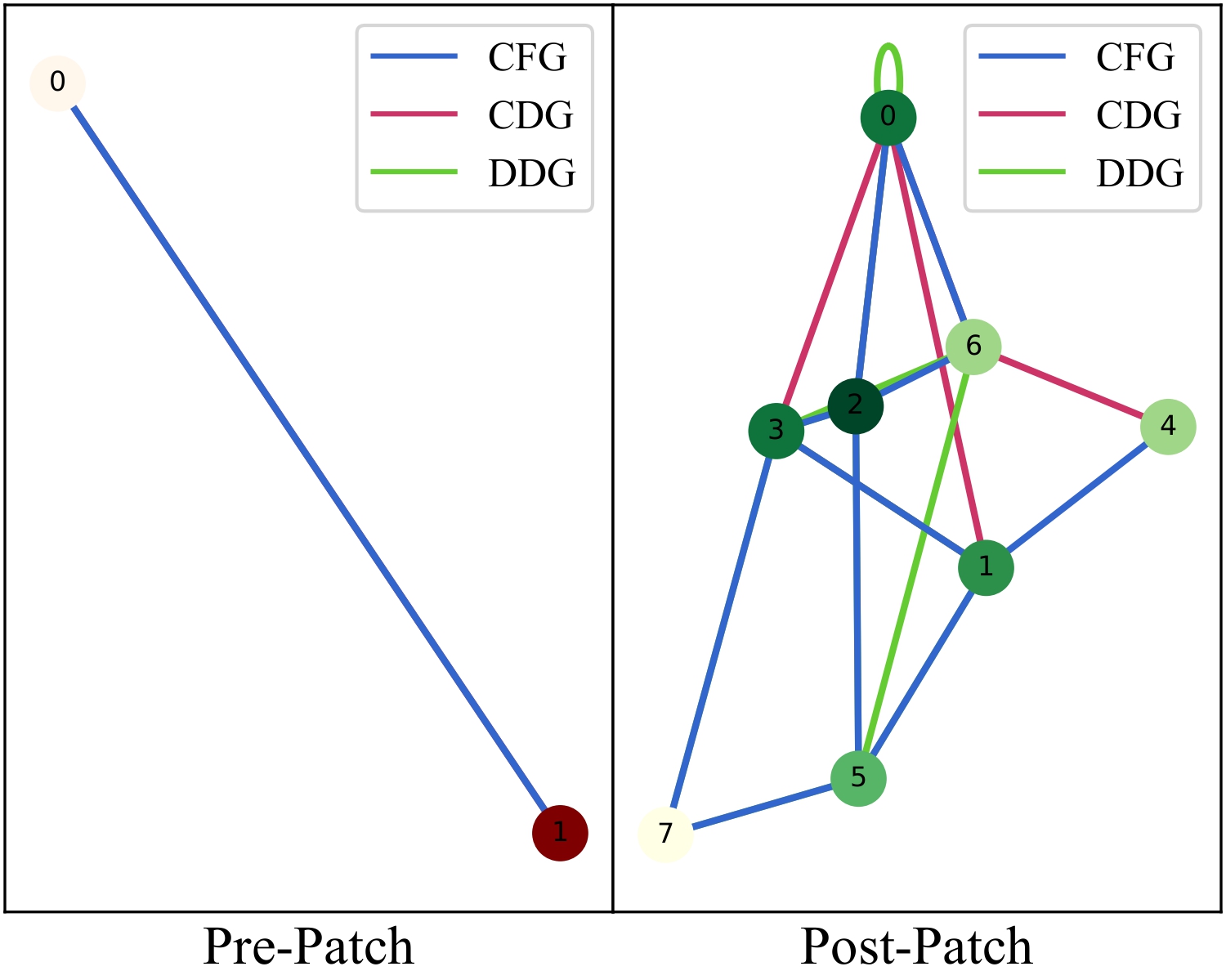}
        }% 
    \vspace{-0.1in}
    \caption{The generated twin graphs using GCC with different optimization levels (CVE-2011-1581).}
    \label{fig:graphs}
    % \vspace{-0.2in}
\end{figure}

After extracting patch-related blocks and building the graph representation, we find that the graph size of security patches is more likely to be smaller than that of non-security ones, as shown in Table~\ref{tab:dataset}. 
This observation is consistent with the findings in previous works~\cite{shu2023graphspd,zhou2021spi}, which show there are usually fewer code changes in the security patches than those in non-security patches.
Besides, we can also confirm that the binary code characteristics, i.e., the graph representations may vary with different compilers and optimization levels, even for the same patch. One example (CVE-2011-1581) is shown in Figure~\ref{fig:graphs}.

\vspace{0.05in}
\noindent\textbf{Embedding Generation.}
The BERT-based embedding model is implemented using \emph{pytorch} 1.13 and \emph{BERT-pythorch} 0.0.1a4.
We set the model structure with 12 encoder layers, each of which contains 8 attention headers.
To cooperate with the input size of the subsequent graph model, we also resize the embedding dimension of the BERT model to 128.
% 我们发现已经有在跨编译器和优化级的binary数据集上预训练训练的BERT模型，因此我们直接采用此结构。
We adopt a BERT model pre-trained on binary datasets with various compilers and optimization levels~\cite{pretrained_model}.
% the pre-trained BERT model on cross-compiler and optimization-level binary datasets, so we directly adopt this structure~\cite{pretrained_model}.
Based on the pre-trained model, we fine-tune the model to fit our dataset in this paper, which is compiled from the Linux kernel.
The BERT-based model generates the node embedding, while the edge embedding is a vector composed of the CFG, CDG, and DDG connections.
Finally, we integrate the node and edge embeddings together into a unified numeric representation.

\vspace{0.05in}
\noindent\textbf{Graph Learning.}
% The embedding model and graph learning model都是基于pytorch包实现，其中
The graph learning program is implemented with Python 3.9.16 while the neural network model is designed with \emph{pytorch} 1.13 and \emph{PyG} 2.2. 
We also design a \emph{TwinGraph} structure to store and process the pre-patch and post-patch graphs at the same time.
In the training phase, the batch size is set to be 128.
We use Adam optimizer with $\beta_1$ of 0.9, $\beta_2$ of 0.99, and a learning rate of 0.001.
To prevent the overfitting issue in graph classification, we set a large dropout ratio of 0.5 in the training phase.
We also use the cross-entropy loss for the gradient descent.
The training set contains 20,031 graph pairs and the testing set contains 5,008 graph pairs; i.e., the training ratio is set to 0.8.
The training set and testing set do not contain overlapping commits.
The max iteration epoch is 1,000 while the model loss becomes stable.

%% file: 05.evaluation.tex
\section{Evaluation}
In this section, we evaluate the performance of \TN{} on our binary patch benchmark. 
We compare the accuracy of \TN{} with the state-of-art works on the security patch detection task. 
Meanwhile, we evaluate the robustness of \TN{} under different compilation configurations, i.e., compilers and optimization levels.

% In this section, we systematically evaluate the accuracy, effectiveness and scalability of \TN{}.

\subsection{Experiment Setup}
\vspace{0.05in}
\noindent\textbf{Runtime Environments.} 
The dataset generation, patch data preprocessing, graph extraction, and embedding generation are deployed in a Linux server with Intel Xeon E5-2650 @ 2.30 GHz and 512 GB memory, running Ubuntu 20.04 LTS.
The GCN-based identification model is carried out in the Ubuntu 22.04 LTS environment running in Intel Xeon Gold 5122 with 3.60 GHz CPU and 64 GB RAM.
The model training is conducted by one NVIDIA RTX 2080 Ti GPU of 11 GB memory with CUDA 11.7.

% \noindent\textbf{comparison Objects.}

\vspace{0.05in}
\noindent\textbf{Evaluation Metrics.} Security patch detection is a classification task, so we use both general metrics and specific
metrics to evaluate its effectiveness and practicality. 
General metrics, including accuracy and F1-score, are used to evaluate the overall performance of the classification model. 
Specific metrics are used to evaluate the practicality of the detection system, including the false-negative rate (FNR) and false-positive rate (FPR).

\subsection{Performance on Security Patch Detection}

We conduct a series of experiments to answer the following questions.
\begin{itemize}
    \item {\bf RQ1:} Does \TN{} outperform the existing state-of-the-art works? 
    \item {\bf RQ2:} Is \TN{} robust on binary patches across different compilers?
    \item {\bf RQ3:} Is \TN{} robust on binary patches across different optimization levels?
    % \item RQ3: Can \TN{} identify security patch within other repositories?
    % \item RQ4: Can \TN{} identify security patches within real-world Linux kernel based Android images?
    % \item RQ5: Is the embedding model and CPG representation better than existing representations?
\end{itemize}

\begin{table}[t]
\begin{center}
\caption{The performance comparison of \TN{} and two baseline models for security patch detection.}
\vspace{-0.05in}
\label{tab_exp_overall}
% \label{tab_exp_twin}
% \renewcommand{\arraystretch}{1.}
\resizebox{\linewidth}{!}{
\begin{tabular}{c|c|p{0.46in}<{\centering}|p{0.43in}<{\centering}|p{0.37in}<{\centering}|p{0.37in}<{\centering}}
\toprule
\multirow{2}{*}{\bf Model} & \multirow{2}{*}{\bf Dataset} & \multicolumn{2}{c|}{\bf General Metric} & \multicolumn{2}{c}{\bf Specific Metric}   \\ 
\cline{3-6} 
{} & {} & {Accuracy} & {F1-score} & {FN rate} & {FP rate} \\ \midrule
PatchRNN~\cite{wang2021patchrnn}  & Source  & {70.86\%}  & 0.379   & {72.97\%} & 7.70\% \\ 
\midrule
GraphSPD~\cite{shu2023graphspd}   & Source  & {85.28\%}  & 0.557   & {56.51\%} & 5.05\%  \\ 
\midrule
\TN{}                             & Binary  & {80.77\%}  & 0.759   & {29.15\%} & 11.82\% \\ 
\bottomrule
\end{tabular}
}
\end{center}
\vspace{-0.05in}
\end{table}

\vspace{0.05in}
\noindent\textbf{RQ1: Does \TN{} outperform the existing state-of-the-art works?}
As shown in Table~\ref{tab_exp_overall}, the \TN{} system can achieve up to 80.77\% with an F1-score of 0.759 on the binary patch dataset.
We compare \TN{} directly with the state-of-the-art source code level detection systems PatchRNN~\cite{wang2021patchrnn} and GraphSPD~\cite{shu2023graphspd}.
Note that, we retrained PatchRNN and GraphSPD based on the Linux kernel part of PatchDB~\cite{wang2021PatchDB}.
Therefore, the experimental results in Table~\ref{tab_exp_overall} are comparable. 

We can observe that the accuracy of \TN{} is slightly inferior to that of GraphSPD, yet it surpasses PatchRNN.
% We can find the performance of \TN{} is better than both two baselines at the source code level.
Specifically, the \TN{} outperforms PatchRNN by 9.91\% of accuracy and 0.380 of F1-score.
Due to the interference caused by the compilation process, the accuracy of \TN{} is 4.51\% lower than GraphSPD.
However, the \TN{} model achieves a higher F1-score compared to state-of-the-art graph models at the source code level, with an improvement of 0.202.
The much higher F1-score indicates that \TN{} has better precision and recall performance.

To investigate such a performance improvement, we can further compare \TN{} with PatchRNN/GraphSPD from the perspective of feature construction and model structure.
PatchRNN adopts a sequential embedding model, similar to word2vec~\cite{alon2019code2vec}, to convert the code changes in patches into an embedding representation and uses the twin-structured RNN model as a classifier.
Similar to \TN{}, GraphSPD constructs the CPGs of both pre-patch and post-patch code; however, GraphSPD merges them into one graph and does not use the sequential embedding model to extract the code semantics. 
Also, GraphSPD utilizes the GCN to learn the merged patch graph representation and to identify the security patches.
In contrast, \TN{} has the advantages of both methods.
From the perspective of feature representation, \TN{} uses the BERT-based embeddings as node attributes in the CPG representation. In addition, \TN{} adopts the \emph{TwinGraph} structure to reserve the features of both pre-patch and post-patch graphs.
From the perspective of model structure, \TN{} adopts the siamese network architecture (same as PatchRNN) to support the pre-patch and post-patch inputs simultaneously; however, each branch of siamese network leverages the GCN structure due to the graph representation.
From the experimental results, the feature fusion and the \emph{TwinGraph} structure play a critical role in \TN{} by capturing more enriched syntax and semantic features.
Although the security patch detection in binary code is more complex than that in source code, 
\TN{} still achieves considerable accuracy and F1-score.

\begin{table}[t]
\begin{center}
\caption{The performance of \TN{} across different compilers and optimization levels.}
\vspace{-0.05in}
\label{tab_exp_compilation}
\renewcommand{\arraystretch}{1.}
\resizebox{\linewidth}{!}{
\begin{tabular}{c|c|p{0.46in}<{\centering}|p{0.43in}<{\centering}|p{0.37in}<{\centering}|p{0.37in}<{\centering}}
\toprule
\multicolumn{2}{c|}{\multirow{2}{*}{\bf Dataset}}  & \multicolumn{2}{c|}{\bf General Metrics} & \multicolumn{2}{c}{\bf Specific Metrics} \\ 
\cline{3-6} 
\multicolumn{2}{c|}{}           & {Accuracy} & {F1-score} & {FN rate} & {FP rate} \\ 
\midrule\midrule
%---------------------------------------------------
\multirow{7}{*}{\bf GCC}     & \texttt{O0}    & {82.2\%}     & {0.78}    & {30.4\%}     & {7.6\%}    \\ \cline{2-6}
{}                       & \texttt{O1}    & {68.5\%}     & {0.62}    & {43.2\%}     & {22.0\%}   \\ \cline{2-6} 
{}                       & \texttt{O2}    & {83.0\%}     & {0.78}    & {26.1\%}     & {10.6\%}   \\ \cline{2-6} 
{}                       & \texttt{O3}    & {81.9\%}     & {0.78}    & {27.1\%}     & {10.6\%}   \\ \cline{2-6} 
{}                       & \texttt{Os}    & {66.3\%}     & {0.54}    & {49.7\%}     & {22.9\%}   \\ \cline{2-6} 
{}                       & Total & {76.4\%}     & {0.70}    & {36.0\%}     & {14.1\%}   \\ 
\midrule\midrule
%---------------------------------------------------
\multirow{7}{*}{\bf Clang}   & \texttt{O0}    & {89.1\%}     & {0.88}    & {14.8\%}     & {7.5\%}    \\ \cline{2-6} 
{}                       & \texttt{O1}    & {86.3\%}     & {0.83}    & {20.0\%}     & {9.3\%}    \\ \cline{2-6} 
{}                       & \texttt{O2}    & {94.6\%}     & {0.94}    & {9.9\%}      & {1.6\%}    \\ \cline{2-6} 
{}                       & \texttt{O3}    & {95.1\%}     & {0.95}    & {7.5\%}      & {2.5\%}    \\ \cline{2-6} 
{}                       & \texttt{Os}    & {90.5\%}     & {0.89}    & {16.3\%}     & {4.2\%}    \\ \cline{2-6} 
{}                       & Total & {90.3\%}     & {0.88}    & {14.7\%}     & {5.7\%}    \\ 
\bottomrule
%---------------------------------------------------
\end{tabular}
}
\end{center}
\vspace{-0.05in}
\end{table}

\vspace{0.05in}
\noindent\textbf{RQ2: Is \TN{} robust on binary patches across different compilers?}
To better understand the impact of compilation on security patch detection, we show the performance of \TN{} under various compilers and optimization levels in Table~\ref{tab_exp_compilation}.
From the results, we notice that \TN{} performs much better on the  patches compiled by Clang than those compiled by GCC, no matter under which optimization level.
% , both in aggregated and subdivided results. 
Specifically, the overall accuracy of \TN{} under Clang is 13.9\% higher than that under GGC; meanwhile, the F1-score under Clang is 0.18 higher than that under GCC.

\begin{figure}[h]
    \centering
    \includegraphics[width=3.2in]{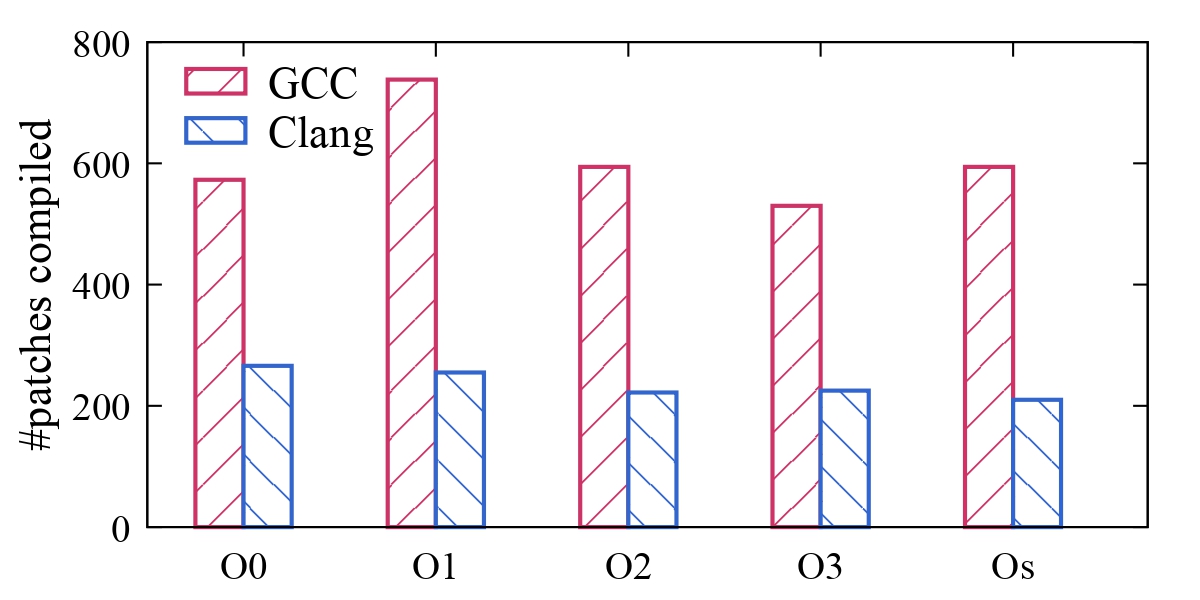}
    \vspace{-0.05in}
    \caption{The number of successfully compiled binary patches across compilers and optimization levels.}
    \label{fig:compiler_opt_success}
    \vspace{-0.05in}
\end{figure}

Another important observation is that the number of patches successfully compiled by Clang is far less than that successfully compiled by GCC, due to the high failure rate of compilation.
As shown in Figure~\ref{fig:compiler_opt_success}, the ratio of patches compiled by Clang and GCC is 1:2.7. 
% Surprisingly, we found that due to the high failure rate of compilation, the number of patches successfully compiled by Clang is far less than the number of successful patches compiled by GCC, reaching a ratio of 1:2.7, as shown in Figure~\ref{fig:compiler_opt_success}.
We maintain the same sample ratio for Clang and GCC in both training and validation phases, which means \TN{} trains on fewer patches compiled by Clang but learns more salient semantics for better recognition performance.

\begin{figure}[t]
    %\vspace{-0.1in}
    \centering
    \subfloat[\#nodes of graphs w/ GCC.]
    {\includegraphics[width=3.3in]{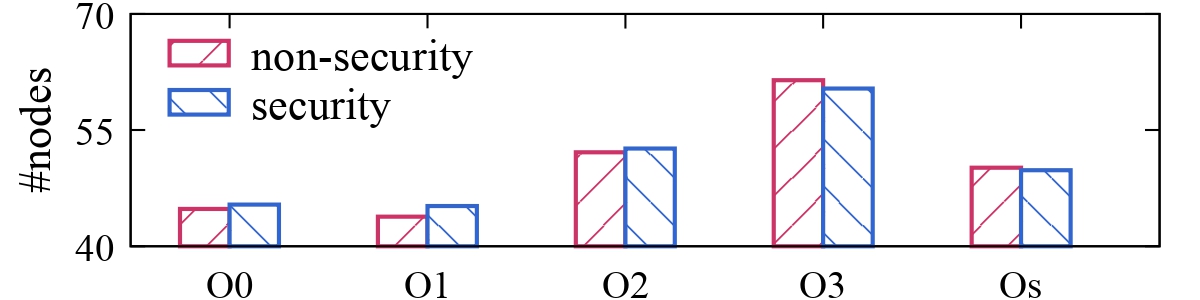}}
    \hfil  
    \vspace{-0.15in}
    \subfloat[\#nodes of graphs w/ Clang.]{\includegraphics[width=3.3in]{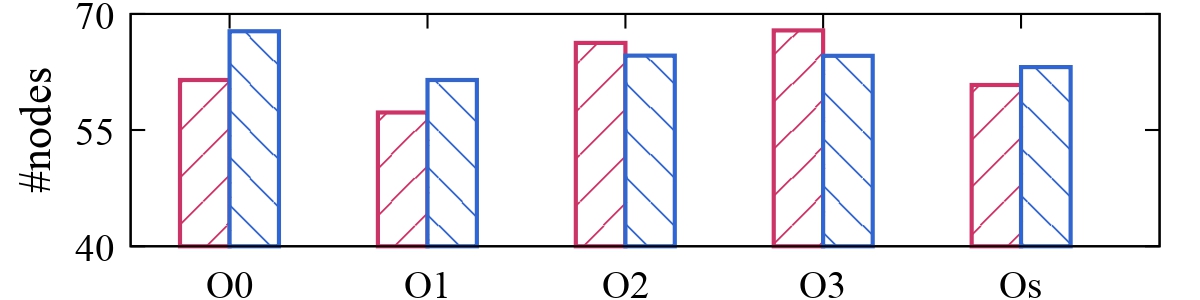}}
    \hfil
    \vspace{-0.15in}
    \subfloat[\#edges of graphs w/ GCC.]
    {\includegraphics[width=3.3in]{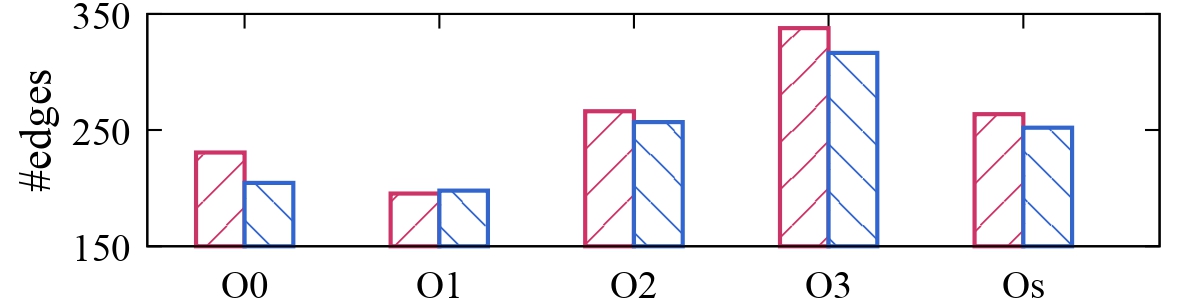}}
    \hfil  
    \vspace{-0.15in}
    \subfloat[\#edges of graphs w/ Clang.]{\includegraphics[width=3.3in]{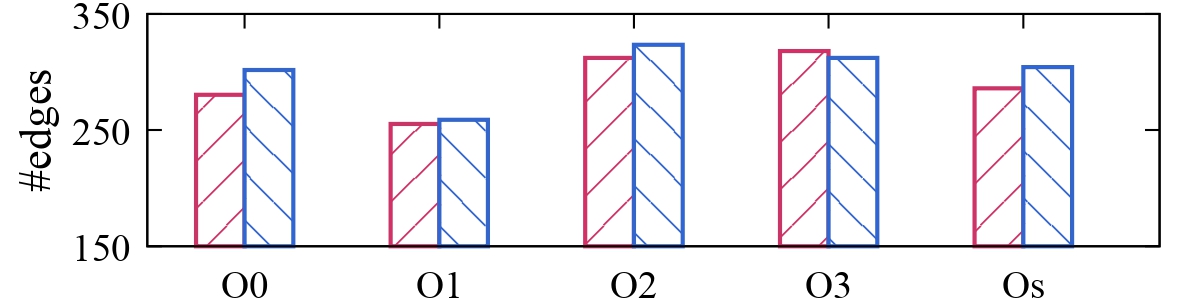}}
    % \hfil
    \vspace{-0.05in}
    \caption{The graph size of patches with different compilers and optimization levels.}
    \label{fig_graphsize}
    \vspace{-0.1in}
\end{figure}

Such a counter-intuitive result is caused by the inherent mechanism of Clang and the code analysis toolchain (i.e., the LLVM toolchain).
On the one hand, the binary patches will be decomposed into more basic blocks due to the compilation mechanism of Clang, hence containing more data dependencies and control dependencies.
% allows binary patches to be decomposed into more basic blocks, 
% and 
We analyze the number of nodes and edges in the CPGs extracted by \TN{} over different compilation configurations.
% from the binary patch in different configurations. 
In Figure~\ref{fig_graphsize}, we count the average number of nodes and edges in both the pre-patch and post-patch graphs.
% in the graph representation of pre-patch and post-patch binaries for each patch.
We find that \TN{} can extract more nodes and edges for the patches compiled by Clang. 
Also, with the compilation of Clang, the graph size difference between security patches and non-security ones is more significant. 
This means, from the perspective of graph representation, the patches compiled by Clang are more distinguishable that those  compiled by GCC.
% This means that, even without relying on the classification model, we can manually find that the patch graph representation compiled from Clang is more distinguishable than GCC's.

On the other hand, we notice that compared with the patches compiled by Clang, the patches compiled by GCC may suffer from the imprecise issue during the disassembly, basic block extraction, and CPG construction.
% the disassembly, basic block extraction, and CPG construction of the GCC-compiled patch might be less accurate than the Clang-compiled patch.
This is because the analysis tools used in this paper are all based on the LLVM toolchain (as mentioned in Section~\ref{Sec: dataset}).
Because Clang belongs to the LLVM toolchain, \TN{} has better compatibility over the patches compiled by Clang.
% Clang belongs to the LLVM toolchain, which has better compatibility. 
% While they may be less accurate on GCC-compiled patches. 
Previous works~\cite{he2022binprov,andriesse2016depth} also introduce the unreliability of reverse engineering tools.

\vspace{0.05in}
\noindent\textbf{RQ3: Is \TN{} robust on binary patches across different optimization levels?}
Table~\ref{tab_exp_compilation} shows the performance of \TN{} over different optimization levels.
% patches compiled with 
% Regarding the impact of optimization levels, we also noticed more detailed changes in Table~\ref{tab_exp_compilation}.
Compiled by Clang, \TN{} can achieve better accuracy and F1-score on the patches with higher optimization levels (e.g., \texttt{O2} and \texttt{O3}); however, the performance of \TN{} on low optimization levels (e.g., \texttt{O0} and \texttt{O1}) is relatively poor, with the accuracy lower than 90\% and an F1-score of 0.88.
Also, \TN{} can achieve moderate performance on the patches optimized with \texttt{Os}.
Nevertheless, the performance gap of \TN{} over different optimization levels does not exceed 10\%.
Compared to Clang, the performance over different optimization levels maintains a similar trend on GCC-compiled patches. 
However, we also notice that the performance of \texttt{Os}-optimized patches drops dramatically to 66.3\% if compiled by GCC.
In this case, the performance gap reaches 16.7\%, compared with both high and low optimization levels.

To investigate the performance trend across optimization levels, we first observe the graph features extracted by \TN{} under different optimization levels.
As shown in Figure~\ref{fig_graphsize}, we notice that \TN{} can extract more nodes and edges for the patches with high optimization levels, while the size of generated graphs with low optimization levels will be relatively small. 
Among different optimization levels, \TN{} extracts the fewest nodes and edges for the \texttt{O1}-optimized patches. 
% Among them, \TN{} extracted the fewest number of nodes and edges from the O1-optimized patch. 
The distribution is consistent with the general performance trend of \texttt{O0}, \texttt{O1}, \texttt{O2}, and \texttt{O3} for both Clang-compiled and GCC-compiled patches. 
Therefore, we can conclude that the higher the optimization level, the better the identification performance of \TN{}.
Moreover, after comparing the number of nodes and edges for security patches and non-security patches, we find that the patches compiled by GCC with \texttt{Os} have the most subtle differences (as shown in Figure~\ref{fig_graphsize}(a) and \ref{fig_graphsize}(c)). 
This also explains why the performance of \TN{} drops drastically under this scenario.

Second, we find that GCC and Clang support different sets of optimization flags. For the GCC, \texttt{Os} selects optimization options in \texttt{O2} that do not increase the object file size; while for Clang, \texttt{Os} is a combination of \texttt{O2} with extra optimizations to reduce code size.
Therefore, compared with \texttt{O2}, there are even fewer optimization items in \texttt{Os} if compiled by GCC.
This is another possible reason for the performance of \TN{} in this case.
In practice, \texttt{O2} is the default optimization level used in most scenarios, while \texttt{Os} is merely used in compilation environments with tight memory resources. 
Therefore, the performance of \TN{} under the GCC compiler with \texttt{Os} optimization is acceptable.

\subsection{False Positive and False Negative}
In the security patch detection task, a false positive (FP) means that a real non-security patch is misidentified as a security patch. 
A false negative (FN) represents that a real security patch was missed out.
From Table~\ref{tab_exp_overall}, we can observe that \TN{} achieves the best FNR of 29.15\%, which is only half of the FNR in these two baseline approaches.
Also, the FPR (11.82\%) of \TN{} is worse than that of PatchRNN (7.70\%) and that of GraphSPD (5.05\%). 
That means the users may get twice as many false security patch alerts when using \TN{} to detect binary patches.
However, the FPR is acceptable considering the extreme imbalance between non-security and security patches in practice. 
The security patches only account for 6-10\% in open source software (OSS)~\cite{wang2021PatchDB}.
Therefore, both FPR and FNR of \TN{} are better than those of the previous methods in general, which is consistent with the results in the F1-score.

Next, we compare the specific performance (FNR and FPR) of \TN{} across compilers and optimization levels.
We observe that the trend of FPR and FNR of \TN{} is consistent with that of F1-score.
It is worth noting that even in the worst case (under GCC with \texttt{Os} optimization), our \TN{} still outperforms previous methods.
Considering that the comparison in this paper is between source code patches and binary patches, the actual performance of \TN{} can even advance more against the previous methods.

%\textbf{Accuracy on new Linux kernel version.} The performance on Linux kernel 5.9 or 5.10. 

% \textbf{Accuracy on popular repositories.}
% The performance on popular repositories, such as openssl, ffmpeg, php and wireshark, etc.

% \textbf{Performance on Android images.}
% The performance on Android images.

%% file: 06.discussion.tex
\vspace{-0.05in}
\section{Discussion}
% 我们的方法具有极强的鲁棒性，在识别由不同编译配置的二进制补丁上
% 作为第一个区分二进制安全补丁的系统，BinGo的效果甚至超过了现有最新的在源码层面的工作。
% 第二，BinGo is robust in identifying binary patches composed of different compilation configurations.
% 这意味着，BinGo成功消除了编译带来的影响。
The evaluation results demonstrate that \TN{} can effectively identify the security patches in binary code.
Although the composition of binary code changes with the compiler and optimization level, \TN{} is robust to eliminate such noisy impact caused by compilation.
% As the first system to distinguish binary security patches, 
\TN{} is able to identify hidden security patches released by the software vendors or from any binary code diffing between neighboring software versions.
% outperforms even the state-of-the-art approaches at the source level.
% Second, \TN{} is robust in identifying binary patches compiled under different configurations.
% This means that BinGo has successfully eliminated the impact caused by compilation.
% 我们的工作可以帮助开发者或者IT操作员发现存在安全补丁的binary，并有选择地进行更新，从而维持功能的稳定性和安全性。
% 另一方面，我们的调查发现，58.55%（469/801）的安全补丁都出现同一个函数内。如果我们在binary中修复漏洞，我们的工作可以帮助定位要修改的函数，开发者可以仅替换相关函数。这降低了人工修复的复杂度。
\TN{} can help developers or IT operators discover binary with security patches, and update them selectively, so as to maintain the stability and security of functions.
Besides, our survey found that 58.55\% (469/801) of security patches appear in the same function. If we fix a bug in the binary, \TN{} can locate the function to be modified, and the developer can replace only the relevant function. This reduces the complexity of manual repair.

% 同时，我们也注意到BinGo仍然有一些局限性，因此一些进一步工作可以考虑。
% 首先，BinGo实际上并未考虑如何分割两个补丁共存于一个binary版本。现有工作已经在源码层面，比如Java,实现了Patch的分割，然而在Binary领域暂无同类工作。
% 其次，BinGo在构造CPG表达时没有考虑被调函数的语义。即，patch如果涉及到新增函数调用时，CPG的构造只考虑调用者函数内的基本块，而不包括被调用者内的代码。
% 但是，如果纳入所有的被调用的程序，CPG会太大以至于难以分析。因此，在进一步工作中，还需要进一步权衡图表示的构造。
% 此外，安全补丁还可以进一步细分。这可以消除false negatives. 我们数据集中安全补丁的类别分布并不均匀。因此，在识别少数类别的安全补丁时，false negative概率会偏高。因此我们可以首先区分安全补丁的类别，从而按类别划分数据集训练模型以降低false negative rate.
However, we also notice that BinGo still has some limitations, so corresponding further work can be taken into consideration.
First, \TN{} does not consider a scenario that two patches coexist in one binary version, in which we need to split the tangled patches.
Existing work~\cite{wang2019cora,shen2021smartcommit} has achieved patch untangling at the source code level, but there is no similar work at the binary level so far.
Therefore, we leave the untangling solution for binary patches in our future work.
Second, \TN{} does not consider the semantics of the callee function when constructing the CPG representation. 
That is, if the patch involves a function call, the CPG only involves the basic blocks in the caller function, but not those in the callee. 
However, if all callee procedures are included, the CPG will be too large to be learned. 
Therefore, in further work, the trade-offs between more semantics and compact graphs are required.
Additionally, security patches can be further categorized, which may be helpful to reduce false negative rates. 
In our dataset, the category distribution of security patches is not uniform, e.g., the If-then-else structures account for more than 70\% of security patches.
Accordingly, when identifying security patches of a minority category, the false negative rate may be high. 
% Therefore, we can first distinguish the security patch type before.
With more security patch types, \TN{} can be well-trained to better classify security patches.

% so as to divide the data set training model by category to reduce the false negative rate.

% \textbf{The extensibility of \TN{}} 
% \TN{} can be extended to train the more general model on security patches from other open-source repositories.
% The current \TN{} only consider security patches whose code changes are within functions. 
% \TN{} can be extended to support complicated security patches via considering function call relation.

% 进一步工作可以做patch的分割。
% \textbf{Reducing false negatives and false positives} 
% The incompleteness of disassembly and graph generation of Angr tools.
% 进一步工作还可以做安全补丁的类别的细分。
% Further work can also be done to subdivide the categories of security patches. 
% This can eliminate false negatives. As mentioned in the previous section, the category distribution of security patches in our dataset is not uniform. 
% Therefore, when identifying security patches of a minority category, the probability of false negatives will be high. 
% Therefore, we can first distinguish the categories of security patches, so as to divide the data set training model by category to reduce the false negative rate.
% \textbf{Security patch category} 

%% file: 07.relatedwork.tex
\section{Related Work}
\vspace{0.05in}
\noindent\textbf{Code Representation.}
% 不管是源码还是二进制码，在代码分析中，代码的表达都是第一步骤。
Regardless of whether it is source code or binary code, extracting the code representation is the first step in the program analysis.
The graph representation is the natural way to represent the intrinsic program flow structure~\cite{allamanis2018learning,Learning2020Wenhan,Yamaguchi2014Joern}.
% It is natural to represent code as graphs because of the intrinsic control flow and data flow structures [12], [13], [71]. 
% 通过定义不同的边和节点，
By defining different edges and nodes, graphs can be utilized for diverse program analysis tasks.
For example, Yamaguchi~\etal~\cite{Yamaguchi2014Joern} first defined the code property graph (CPG) by integrating multiple types of graphs to detect the vulnerabilities.
Ji~\etal~\cite{Yuede2021Vestge} encoded multiple statistical features into the attributed control flow graph (ACFG) to identify the compilation provenance.
% automatic source code summarization [72], learning semantic program embeddings [73], and learning compiler optimization tasks [x].
%
Inspired by the popularity of the large language model (LLM), researchers employed self-supervised sequence embedding models~\cite{devlin2018bert, radford2018improving, lewis2019bart} to capture the textual semantics from code~\cite{liu2022commitbart, feng2020codebert, li2021palmtree}. 
% The LM-based embedding model在程序合成和debug的任务上已经取得了不错的表现。
In industry, the LLM-based embedding model has achieved good performance in code generation and debugging tasks, such as the Codex~\cite{codex}, and ChatGPT~\cite{chatGPT}.

\vspace{0.05in}
\noindent\textbf{Patch Analysis.}
% 由于开源和程序复用的发展，补丁分析对软件供应链的安全变得尤为重要
Due to the development of open-source software (OSS) and program reuse, patch analysis has become especially important for the security of the software supply chain~\cite{li2017large,synopsys-report}. 
% 王等人从NVD数据库和github收集了海量的补丁并提供了banchmark用于补丁分析。
Wang~\etal~\cite{wang2021PatchDB} collected massive patch information from the NVD database and GitHub and provided a benchmark for patch analysis.
Li~\etal~\cite{li2017large} empirically studied the syntax structure of security patches, revealing multiple significant behaviors.
% 补丁分析有助于多个下游任务。
Patch analysis facilitates several downstream tasks, such as vulnerability detection and automated program repair.
% 一方面，安全补丁检测是漏洞检测的镜像问题。
On the one hand, security patch detection is the mirror problem of vulnerability detection~\cite{BScout2020,zhang2021investigation,zhang2018precise,PDiff2020Jiang}.
% 检测安全补丁意味着对应的漏洞已经被修复。反之，则说明程序中漏洞依然存在。
Detecting a security patch indicates the corresponding vulnerability has been fixed. 
% On the contrary, it means that the loopholes in the program still exist.
Tian~\etal~\cite{Identifying2012Tian} utilized textual and code features of source code to detect bug-fixing patches in Linux.
Zhang~\etal~\cite{zhang2018precise} proposed the patch presence test for binaries, which maps the patch patterns in source code changes and checks the presence of such patterns in the binary code.
% 另一方面，研究者们分析安全补丁的语法和语义特征，并用于自动地生成补丁。
On the other hand, researchers analyze the pattern and structure of security patches and imitate them to generate patches automatically~\cite{patching2019NDSS, HERA2021}.
% 现有工作从现有代码库中学习安全补丁的模板并将其迁移到新的风险代码中 [][][]。
Aravind~\etal~\cite{spider2020machiry} defined the safe patch that does not disrupt the intended functionality of the program. Such a safe patch can be propagated in the software supply chain. 
Xu~\etal~\cite{hotpatch2020xu} proposed Vulmet which can automatically generate hot patches for Android via learning patch semantics. 
Wang~\etal~\cite{wang2020machine} further used the random forest with extracted patch features to classify security patches into specific vulnerability types. 
% Apart from Learning-based solutions, recent works~\cite{Precisely2020,huang2019using} proposed rule-based methods to generate security patches by instantiating a pre-defined patch template.
% Wu et al. and Huang et al. [46], [47] developed rule-based methods to identify some common types of security patches. 
% Correspondingly, related techniques are also proposed in binary codel level involving binary diffing [52], [53], [54], patch presence test [55], [56], [57], [58], patch identification [59] and automated binary patching [60], [61].

\vspace{0.05in}
\noindent\textbf{Security Patch Identification.}
% 上述所有工作有一个共同的前提，目标补丁一定是安全相关的。区分安全补丁和非安全补丁保证了这一假设。
All the above works have a common premise that the object patch must be security-related.
Distinguishing between security patches and non-security patches can warrant such an assumption.
Researchers constructed syntactic and semantic features from the source code and commit messages, then employed machine learning (ML) or even deep learning (DL) based classifiers to distinguish security patches from other patches~\cite{PatchNet2019Hoang}.
For example, PatchRNN~\cite{wang2021patchrnn} and SPI~\cite{zhou2021spi} identify security patches with RNN models. 
Wang~\etal~\cite{shu2023graphspd} developed graphSPD, which conducted multiple-attribute graph representation based on the commit information and exploited the graph learning model to detect  security patches. 
Regarding binary patch detection, to the best of our knowledge, there is no existing work that focuses on distinguishing security patches from non-security ones. 
Thus, \TN{} is the first approach that focuses on detecting binary security patches.

%% file: 08.conclusion.tex
\vspace{-0.05in}
\section{Conclusion}
In this paper, we propose \TN{}, a new end-to-end binary patch identification system.
\TN{} can be performed in two steps.
\TN{} first leverages the code property graph and LM-based embedding model to encode the semantics of binary patches, 
and then distinguishes the security patches via a Siamese-structured GCN model. 
\TN{} can help users and developers select critical security patches from unknown binary patches to ensure software security and functional stability.
To train and evaluate \TN{}, we also propose an automatic approach to generate a binary patch dataset according to known patch information in the source code. 
We implement a prototype of \TN{} and conduct experiments to evaluate the effectiveness and robustness of \TN{}.
The experimental results show that \TN{} can achieve great performance (80.77\% accuracy and 0.759 F1-score) for identifying security patches.
Binary patches can be harder to detect compared to source code patches due to various compilation configurations.
However, \TN{} outperforms the state-of-art solutions that focus on identifying patches in source code, and exhibits good robustness to patches across compilers and optimization levels.
In addition, the experimental results show that \TN{} has fewer false alarms, which means that \TN{} rarely misses or misidentifies security patches.

% \TN{}可以帮助用户和开发者从混合的更新包中挑选至关重要的安全补丁，从而确保软件的安全性和功能稳定性 
% \TN{} leverages a BERT-based embedding model to learn the contextual semantics in binary code.
% The only feature used in \TN{} is byte sequences, which can be directly extracted from binary code. 
% It can be further extended to use in other downstream tasks, such as vulnerability detection, patch presence, and hot patch generation.
% Based on the benchmark, 
% 尽管二进制补丁会因为不同的编译配置变得复杂，\TN{}在训练和验证时，对不同的编译器和优化级有很好的泛化性。
% Regarding the security patch detection task, 
% 而且表现出很好的鲁棒性对于跨编译器和优化级的补丁
% 此外，实验结果显示\TN{}具有较低假阴性率(6.37%),这意味着\TN{}很少会错过真的安全补丁。
% generalizes well to different compilers and optimization levels during both training and validating.
% We can therefore infer the optimization level more accurately. 
% 关于安全补丁检测任务，二进制补丁会比源码补丁更难以检测，由于多样的编译配置. 然而，\TN{}